\documentclass{elsarticle}

\usepackage[utf8]{inputenc} 
\usepackage[T1]{fontenc}    
\usepackage{setspace}
\usepackage{fullpage}
\usepackage{hyperref}       
\usepackage{url}            
\usepackage{booktabs}       
\usepackage{amsfonts}       
\usepackage{nicefrac}       
\usepackage{microtype}      
\usepackage{amsmath}
\usepackage{longtable}
\usepackage{dcolumn}
\usepackage{graphicx}
\usepackage{color,soul}
\usepackage{lineno}
\usepackage[final]{changes}
\usepackage{scalefnt}

\usepackage{setspace}

\begin{document}

\title{TeaNet: universal neural network interatomic potential inspired by iterative electronic relaxations}

%

\author[1]{
  So Takamoto\corref{cor1}
}
\ead{
  takamoto.so@fml.t.u-tokyo.ac.jp
}
\author[1]{
  Satoshi Izumi
}
\ead{
  izumi@fml.t.u-tokyo.ac.jp
}
\author[2]{
  Ju Li
}
\ead{
  liju@mit.edu
}
\cortext[cor1]{Corresponding author}
\address[1]{
  Department of Mechanical Engineering\\
  The University of Tokyo\\
  7-3-1 Hongo, Bunkyo-ku, Tokyo 113-8656 \\
}
\address[2]{
  Department of Nuclear Science and Engineering and Department of Materials Science and Engineering\\
  Massachusetts Institute of Technology \\
  Cambridge, MA 02139 \\
}

\begin{abstract}
A universal interatomic potential for an arbitrary set of chemical elements is urgently needed in computational materials science. Graph convolution neural network (GCN) has rich expressive power, but previously was mainly employed to transport scalars and vectors, not rank $\ge 2$ tensors.  As classic interatomic potentials were inspired by tight-binding electronic relaxation framework, we want to represent this iterative propagation of rank $\ge 2$ tensor information by GCN. Here we propose an architecture called the tensor embedded atom network (TeaNet) where angular interaction is translated into graph convolution through the incorporation of Euclidean tensors, vectors and scalars. By applying the residual network (ResNet) architecture and training with recurrent GCN weights initialization, a much deeper (16 layers) GCN was constructed, whose flow is similar to an iterative electronic relaxation. Our traning dataset is generated by density functional theory calculation of mostly chemically and structurally randomized configurations.  We demonstrate that arbitrary structures and reactions involving the first 18 elements on the periodic table (H to Ar) can be realized satisfactorily by TeaNet, including C-H molecular structures, metals, amorphous SiO${}_2$, and water, showing surprisingly good performance (energy mean absolute error 19 meV/atom) and robustness for arbitrary chemistries involving elements from H to Ar.
\end{abstract}


\maketitle

\section{Introduction}

A universal interatomic potential for atomistic simulations of
arbitrary chemical species, structures, transformations and reactions
would considerably extend the reach of computational materials.  While
historically we have used simple analytical expressions
\cite{PhysRevB.29.6443, PhysRevB.39.5566, PhysRevB.97.125411}, machine
learning (ML) interatomic potentials \cite{PhysRevLett.98.146401,
  gilmer2017neural, schutt2017quantum, schutt2017schnet,
  bartok2010gaussian} are increasingly invoked to parametrize
interatomic interactions.

Deep neural networks (DNN) have proved to be successful in various ML
tasks when large datasets are provided.
The convolution operation, where identical set of weights are used for
nodes ``belonging'' to different spatial locations, achieves efficient
compression.  The convolutional weight depends on the relative
distance, and not the absolute positions (``translational
invariance'').  This idea of parametrizing interactions by spatial
relationships can be generalized to graphs.  The field of graph
convolution-based neural networks (GCN) has been expanding rapidly
\cite{4700287, bronstein2017geometric, grover2018graphite}, in
particular for molecular systems, where atoms and bonds are
represented by the nodes and edges of the graph. Such network
architectures appear natural to both atomistic and
electronic-structure modelers. Indeed, as all the atoms/ions of the
same chemical type/valence state and isotopic mass are
``indistinguishable particles'' in quantum mechanics, the GCN weights
assigned to atoms/bonds of the same chemical type(s) but different
integer labels $i$ or $j$, where $i,j=1,2,...,N$ is the (arbitrarily)
assigned index of an atom in the simulation, should obviously also be
identical (``permutational invariance''). However, sometimes, like in
multi-identical-Fermion wavefunction, there can be a ``minus sign''
issue.  Such ``minus sign'' can show up in some bond-centered
quantities, e.g. if ${\bf x}_{ij}\equiv {\bf x}_i - {\bf x}_j$, then
${\bf x}_{ij} = -{\bf x}_{ji}$, and how to store certain
``bond-centered'' quantities thus necessitates the usage of notation
$[ij]$ where the order of $i,j$ in the bracket matters, unlike
$r_{ij}\equiv |{\bf x}_{ij}|$ where the order of $i,j$ does not
matter, for which we use the notation $(ij)$. So we use notation ${\bf
  x}^{[ij]}\equiv {\bf x}_{ij}$ to denote a vector that belongs to
directed edge labeled by $[ij]$, and $r^{(ij)}\equiv r_{ij}$ to denote
a scalar that belongs to undirected edge labeled by $(ij)$, for
``bond-centered'' quantities, that can be scalar (rank-0 tensor),
vector (rank-1 tensor), matrix (rank-2 tensor), etc. Note in this
paper we take ``bond'' to mean $i,j$ pair relations where $r^{(ij)}$
can nanometers, and not necessarily the so-called first nearest
neighbors.

While GCN architecture exploiting translational and permutational
invariances remove the dependence on an arbitrary observation-frame
origin and an arbitrary atomic indexing scheme, how ``rotational
invariance'', that is, how arbitrary observation-frame orientations
affect or not affect certain results, needs to be discussed.  In any
atomistic calculation of the stress tensor, heat flux vector, etc. based on
for instance the Tersoff potential \cite{LiPY98}, or in assembling the
electronic overlap integral and Hamiltonian matrix in the
tight-binding / linear combination of atomic orbitals (LCAO) model
\cite{QianLQWCYHY08}, one has plenty of scalars (rank 0), vectors
(rank 1) and matrices (rank 2) in the data flow of a code.  In an
iterative electronic relaxation or explicit time-dependent density
functional theory (TDDFT) \cite{QianLLY06} calculation, this kind of
tensorial data flow can sometime even carry into the
(pesudo)time-domain.  In all these calculations, the observation-frame
orientation does not really matter, as all physical quantities are
expressed in rank-$M$ tensors $\tilde{T}_{\alpha_1, \alpha_2, ...,
  \alpha_M}$, with tensor transformation law
\begin{equation}
  \tilde{T}_{\alpha'_1, \alpha'_2, ..., \alpha'_M} \;=\; Q_{\alpha'_1
    \alpha_1} Q_{\alpha'_2 \alpha_2} ... Q_{\alpha'_M \alpha_M}
  T_{\alpha_1, \alpha_2, ..., \alpha_M}
\end{equation}
where $\tilde{T}$ is the same physical object read in a different
observation frame, $Q_{\alpha' \alpha}$ is the rotation matrix between
two observation frames, and Einstein summation rule is used. In this
paper we use $\alpha,\beta=1,2,3$ to label Cartesian axes, and
$i,j,k=1,2,...,N$ to label atoms.  Thus, $T_{\alpha}^{[ij]}$ denotes a
rank-1 tensor (vector) that belongs to a bond, or pair of atoms
$[ij]$, where the order matters (directed edge), and
$T_{\alpha\beta}^{(ij)}$ is a rank-2 tensor (matrix) that belongs to
the bond or pair of atoms $(ij)$ where the order does not
matter. Similarly, $T_{\alpha}^{i}$ is a rank-1 tensor (vector) that
belongs to the atom $i$, and $T_{\alpha\beta}^{i}$ is a rank-2 tensor
(matrix) that belongs to the atom $i$.  One could certainly come up
with more complex notations like $T_{\alpha\beta}^{(ijk)}$ where the
permutation orders of $i,j,k$ does not matter, or something like
$T_{\alpha\beta\gamma\delta}^{[(ij),(kl)]}$ where $i,j$ order does not
matter, $k,l$ order does not matter, but $(ij)$ and $(kl)$ order
matters. Generally speaking, in this notation the superscript denotes
the ``owner'' of the tensor whose Cartesian indices are in the
subscript. In this paper we will only be limited to $M\le 2$, and
owners either $i$, $(ij)$, or $[ij]$, as these covers the data types
of most of the legacy codes.  One can thus imagine these kinds of
``tensor-typed'' and ``ownership-stamped'' data flowing in respective
legacy codes to represent interatomic or electronic-structure
interactions.

In addition to the stable molecular structures, currently several GCN
models have also succeeded in reproducing the dynamics of specific
molecules \cite{schutt2017schnet, schutt2017quantum,
  PhysRevLett.120.143001, doi:10.1021/acs.jctc.9b00181,
  doi:10.1021/acs.jpclett.9b02037}. However, a universal IP describing
bond formation, bond breaking and recombination for arbitrary
structures with arbitrary number of elements remains at the
developmental stage. Inspired by the nonlinear iterative data flows in
a DFT calculation in achieving charge-density convergence, we believe
the performance of GCN can be significantly improved by allowing $M\ge
2$ quantities (``tensors'' in ``tensor embedded atom network
(TeaNet)'') to flow in the network, in addition to the $M=0$ (scalars)
and occasional $M=1$ (vectors) quantities that flow in conventional
GCN.

Physically, embedded atom method (EAM) potential incorporates the
concept of electron density of metal, while Tersoff-type and modified
embedded atom method (MEAM) potentials incorporate the concept of bond
order and angular dependence, which can be derived from the
tight-binding approximation of the electronic wave function, using
local combination of (quasi)atomic orbitals \cite{QianLQWCYHY08}.
These IPs have been widely used for simulating extended defects,
mechanical deformation and damage, and phase transitions.  However,
individual potential parameter set is developed to reproduce a certain
systems (e.g. FCC metals, silica, organic molecules, etc.).  In this
paper, we propose a NNIP architecture (see section
\ref{sec:architecture}) that can be considered a superset of MEAM
potentials while mimicking electronic total-energy relaxation
\cite{PayneTAAJ92} in a local orbital (tight-binding) basis
\cite{QianLQWCYHY08,WangLLYH08,WangLYLYH08}. We call this approach the
tensor embedded atom network (TeaNet).  We modify the architecture of
GCN with new components (edge-associated in addition to
node-associated variables) that fully represent the corresponding
physics-based IP. Rank-2 tensors as well as rank-1 vectors are
introduced in the network, so the model can naturally represent
propagation of orientation-dependent Hamiltonian information.  We have
also adopted residual NN architecture, with recurrent parameter model
initially.  Such ResNet architecture and recurrent GCN initialization
to accelerate computations are found to be quite effective in getting
rapid reduction of traning error.

Our method is related to previous NNIP efforts. Embedded Atom Neural
Network Potential (EANN) \cite{doi:10.1021/acs.jpclett.9b02037}
extends the EAM potential using NN. This model combines physics-based
representation (electron density) and NN-based embedding function
$F(\rho)$. This physics-related model provides excellent accuracy for
bulk systems while retaining simplicity. There are several works 
implementing higher-order geometric information into NN architecture
through spherical harmonics representation \cite{kondor2018clebsch,
  anderson2019cormorant, 47487}. The key idea is to use Clebsch-Gordan
coefficients to hold invariances by any rotations in SO$(3)$ group. In
addition, the idea and the theoretical study of using tensor values in
the interatomic potential was investigated in Moment Tensor Potentials
(MTP) \cite{doi:10.1137/15M1054183}.

In section \ref{sec:train_result}, we show the training results of our
model for elements 1-18 (H-Ar) on the periodic table, where random
combination of these elements in mostly highly disordered structures
are used as the training set. We also performed sensitivity analysis and
discussed the importance of the different features of our model.  In
section \ref{sec:applications}, we show the general applicability of
our method to a wide range of materials including chemical reaction
processes. We will demonstrate that our model performs well for liquid
water, amorphous silica as well as simple metals and hydrocarbons.

\section{TeaNet architecture}
\label{sec:architecture}

\begin{figure}[th]
	\centering
	\includegraphics[width=0.9\linewidth,trim=00 00 00 00]{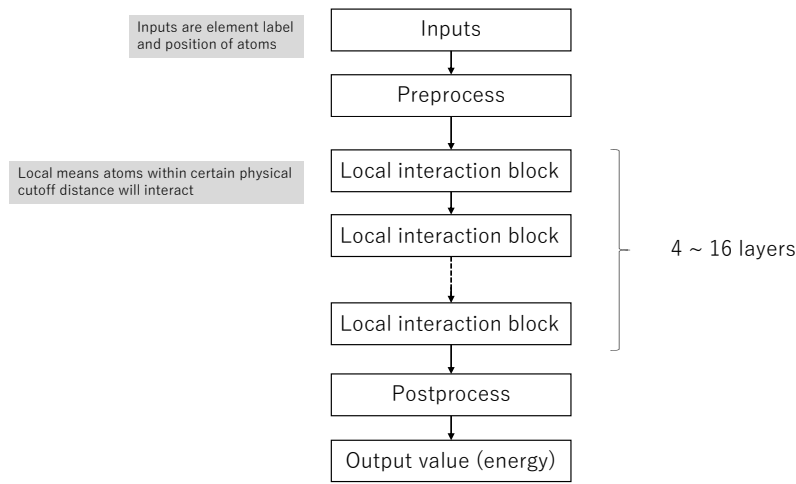}
	\caption{Overview of the TeaNet NNIP.}
	\label{fig:SIOverview}
\end{figure}

We first introduce the notation used in our drawings.  In the
line-drawing figures, values are illustrated as circles. The filled
colors corresponds to the types of the values, where scalar, vector,
and tensor are illustrated as gray, light blue, and blue circles,
respectively.  Operations are illustrated as rectangles. Here, we
write the linear layer as $\mathrm{lin}(x)$, the nonlinear activation
layer as $\mathrm{act}(x)$, the concatenation function as
$\mathrm{con}(x, y, ...)$, vector ${\rm L}2$-norm as
$\mathrm{norm}(x)$, and the cutoff function as $\mathrm{cut}(x)$.  We
use subscript to denote the dimension of stacked variables, for
example ${\rm s}_{128}$ means 128 scalars, ${\rm v}_{32}$ means 32
vectors (total 48 real numbers), and ${\rm t}_{16}$ means 16 matrices,
each $3\times 3$.  The 128, 32, and 16 are called number of channels.

$\mathrm{lin}(x)$ is always applied channel by channel. It is noted
that each $\mathrm{lin}(x)$ appeared in the following equations has
different parameters. It is also noted that those parameters are
learnable network parameters like in ordinary neural networks.

While the output of TeaNet is the total energy of the system (a
scalar), the network is trained to simultaneously compute the atomic
forces, providing useful data for training. The atomic forces are
calculated by a backpropagation process, and so the training process
becomes a double backpropagation.  The molecular dynamics simulation
requires a smooth activation function. In this study, we employed the
integral of the softplus function, which to our knowledge we were
first to propose as an activation function. The integral is calculated
as follows:
\begin{equation}
\begin{split}
  f(x)&\equiv
      \int_{-\infty}^x \log\left(1+\exp\left(t\right)\right)\mathrm{d}t \\
      &=-\mathrm{Li}_2\left(-\exp\left(x\right)\right),
\end{split}
\end{equation}
where $\mathrm{Li}_2$ is a second-order polylogarithm function. This
function approaches $0$ as $x$ tends to $-\infty$ and approaches the
curve of $x^2+C$ at large $x$, where $C$ is a constant. When this
function is applied to the edge arrays, the activation functions are
shifted so that $f(0)$ becomes $0$.  Thus

\begin{equation}
\mathrm{act}(x) \equiv f(x) - f(0).
\end{equation}

\begin{figure}[th]
	\centering
	\includegraphics[width=0.9\linewidth,trim=00 00 00 00]{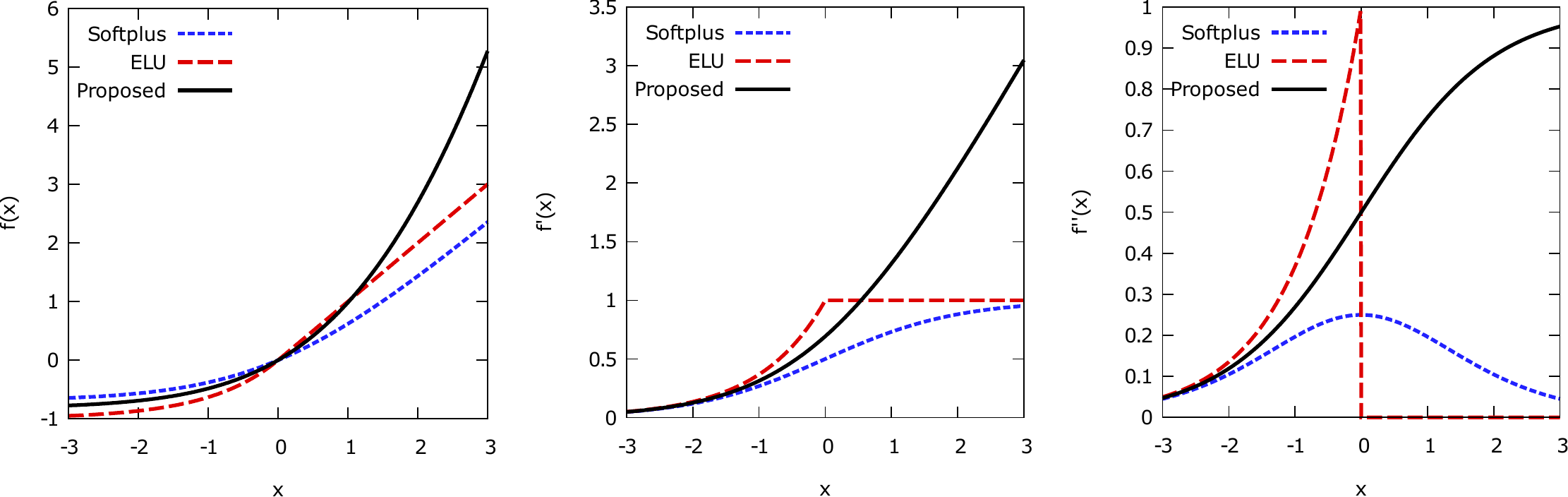}
	\caption{Left: comparison of activation functions. Softplus and ELU ($\alpha=1$) functions\cite{clevert2015fast} are also shown. They are shifted so that $f(0)$ becomes $0$. Middle: derivative of the activation functions. Right: second derivative of activation functions. In softplus and ELU, second gradient value $f''(x)$ vanishes when $x$ is large.}
	\label{fig:activation}
\end{figure}

Using the activation function, we
can train a softplus-type network in the second backpropagation
process. If the polylogarithm function is replaced by the softplus
function, the second backpropagation process results in a sigmoid-type
network.  The function shape is shown in Fig. \ref{fig:activation}.
The effect of this change to the prediction accuracy is presented in
the section \ref{sec:train_result}.

The cutoff function $\mathrm{cut}(x)$ is a smoothly decaying function.
In this work, we use the same function as
$\mathrm{act}(x)$ shifted by linear function,

\begin{equation}
\mathrm{cut}(x) \equiv \mathrm{act}(\mathrm{lin}(x)) + (c_1 x + c_0)
\end{equation}

where $c_0$ and $c_1$ (linear function part) are set to satisfy
$\mathrm{cut}(x)$ and its derivative are zero when $x$ equals to the
cutoff distance.

In TeaNet architecture, the inputs are the list of element label and
the list of position of atom. Other predefined information such as
bonding or atomic charges are not required.  The output value is
single scalar value, which corresponds to the energy. The force of the
atoms are calculated using normal back propagation. There are three
parts in TeaNet. The first part is preprocess. It receives the input
values and creates various values which is used for the graph
convolution layers. The second part is the internal graph convolution
layers which we call local interaction block. The input values and
output values of the single layer have the same shapes. Therefore we
can stack the layers by arbitrary numbers. The last part is
postprocess part, which receives the ouput values of the graph
convolution layer and output single scalar value.

\subsection{Preprocess and postprocessing}
\label{sec:specification}

\begin{figure}[th]
	\centering
	\includegraphics[width=0.9\linewidth,trim=00 00 00 00]{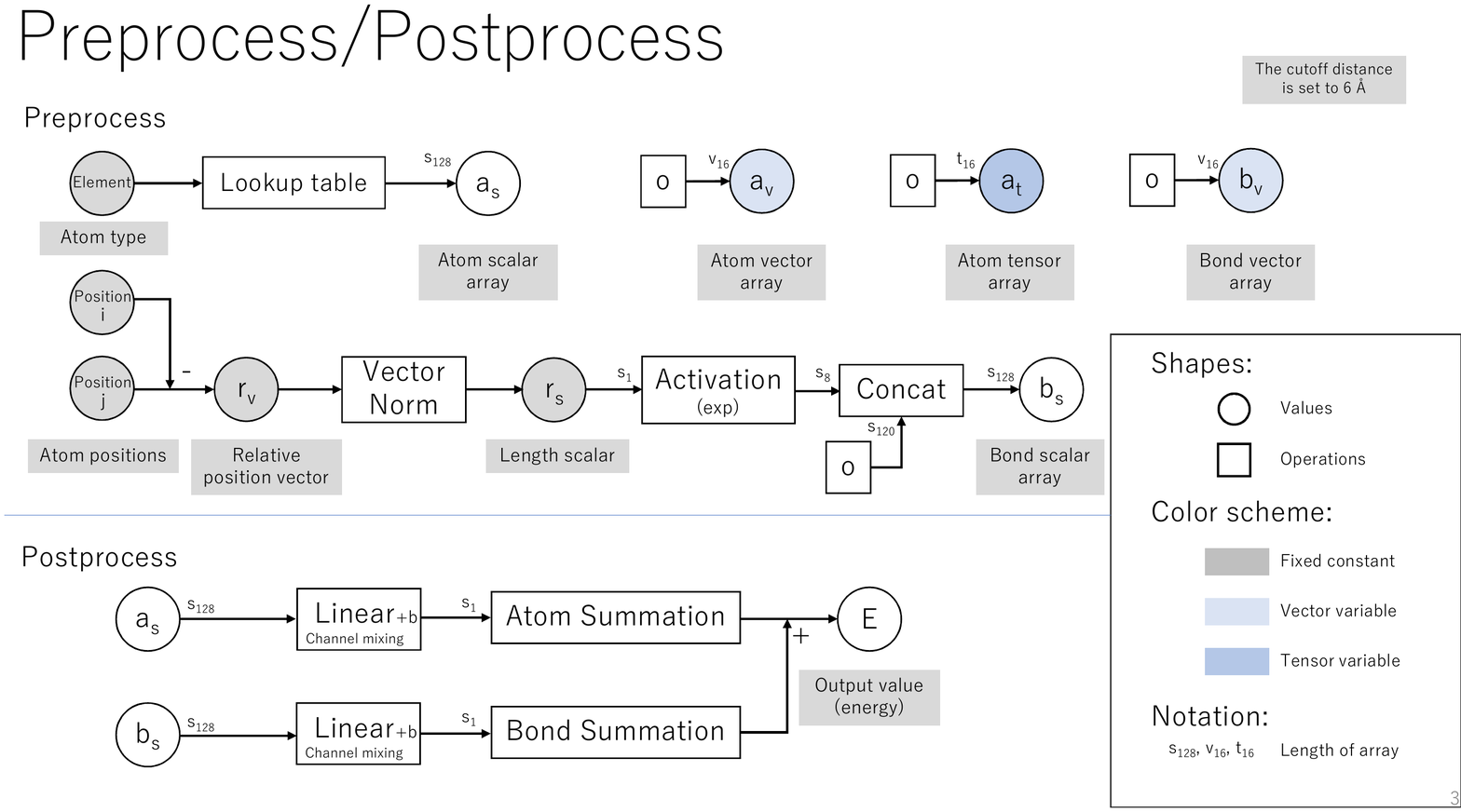}
	\caption{Preprocess and postprocessing}
	\label{fig:SIPreprocessPostprocess}
\end{figure}

\subsubsection{Preprocessing}

Here, we use the character $a$ as atom-related values (corresponding
nodes) and $b$ as bond-related values (corresponding bonds).

Bonds are counted only for pair of atoms whose distance is smaller
than the cutoff distance. In this paper, the cutoff distance is set to
be 6 \mbox{\normalfont\AA}.

There are three types of values for atom-related values which are
scalar, vector, and rank-2 tensor. We use the symbols $a_s$, $a_v$,
and $a_t$ for them. It is noted that each types of values have
multiple channels. For example, in this paper, we use 256 dimensions
(usually called channels in neural network context) for scalar value
and 16 dimensions for both vector and rank-2 tensor value. Therefore,
if the number of atoms in the system is 64 and the number of dimension
of the space is 3, the shapes of $a_s$, $a_v$, and $a_t$ are $64\times
128$, $64\times 3\times 16$, and $64\times 3\times 3\times 16$,
respectively.

Bonds have scalar and vector values. We use the symbols $b_s$, $b_v$
as well. In addition, two special constant values for bond-related
values are also introduced. One is relative position vector $r_v$. It
is defined by the difference of the position of two corresponding
atoms. Another one is bond length $r_s$, which can be calculated by
the l2-norm of $r_v$. It should be noted that the sign of $r_v$
depends on the order of corresponding two atoms, which is described as "minus sign" issue at the introduction section.
Careful consideration is required to use $r_v$ in the following calculations since the
output value should not depend on the order of atoms. We use the
character $i$ and $j$ for the label of those two atoms.

Atom scalar $a_s$ is initialized by look-up table. To imitate the
occupancy of electron orbitals, the values corresponding to the atomic
number are divided by 2 and packed by 1 from the top of the array. The
list is shown in table \ref{tab:input_ns}. The remaining channels are
set to zero. Atom vector $a_v$ and rank-2 tensor $a_t$ are initialized
by zero.

Bond scalar $b_s$ are initialized by Eq. \ref{eq:cut1},
\begin{equation}
b_s=\exp{\left\{-\mathrm{lin}\left(r_s\right)\right\}}+\left(c'_1r_s+c'_0\right), \label{eq:cut1}
\end{equation}
where $c'_0$ and $c'_1$ (linear function part) are set to satisfy $b_s$
and its derivative with respect to $r_s$ are zero when $r_s$
equals to the cutoff distance. Eq. (\ref{eq:cut1}) is expected to
behave like the distance term of the Morse-style IP.

Bond vector $b_v$ is also initialized by zero. Unlike $r_v$, we make
$b_v$ does not depend on the order of atom $i$ and $j$. The example of
physical value corresponding $b_v$ is local electric dipole.

To reiterate, in the preprocess part, $a_s$, $a_v$, $a_t$, $b_s$,
$b_v$, $r_s$, and $r_v$ are initialized.

\begin{table}[th]
	\caption{List of the input values of ${\bf n}_s$.}
	\label{tab:input_ns}
	\centering
	\begin{tabular}{ll}
		\toprule
		\multicolumn{1}{c}{Element} & \multicolumn{1}{c}{${\bf n}_s$}  \\ \midrule
		H & $\left[0.5, 0, 0, 0, 0, 0, 0, 0, 0\right]$ \\
		He & $\left[1, 0, 0, 0, 0, 0, 0, 0, 0\right]$ \\
		Li & $\left[1, 0.5, 0, 0, 0, 0, 0, 0, 0\right]$ \\
		Be & $\left[1, 1, 0, 0, 0, 0, 0, 0, 0\right]$ \\
		B & $\left[1, 1, 0.5, 0, 0, 0, 0, 0, 0\right]$ \\
		C & $\left[1, 1, 1, 0, 0, 0, 0, 0, 0\right]$ \\
		N & $\left[1, 1, 1, 0.5, 0, 0, 0, 0, 0\right]$ \\
		O & $\left[1, 1, 1, 1, 0, 0, 0, 0, 0\right]$ \\
		F & $\left[1, 1, 1, 1, 0.5, 0, 0, 0, 0\right]$ \\
		Ne & $\left[1, 1, 1, 1, 1, 0, 0, 0, 0\right]$ \\
		Na & $\left[1, 1, 1, 1, 1, 0.5, 0, 0, 0\right]$ \\
		Mg & $\left[1, 1, 1, 1, 1, 1, 0, 0, 0\right]$ \\
		Al & $\left[1, 1, 1, 1, 1, 1, 0.5, 0, 0\right]$ \\
		Si & $\left[1, 1, 1, 1, 1, 1, 1, 0, 0\right]$ \\
		P & $\left[1, 1, 1, 1, 1, 1, 1, 0.5, 0\right]$ \\
		S & $\left[1, 1, 1, 1, 1, 1, 1, 1, 0\right]$ \\
		Cl & $\left[1, 1, 1, 1, 1, 1, 1, 1, 0.5\right]$ \\
		Ar & $\left[1, 1, 1, 1, 1, 1, 1, 1, 1\right]$ \\
		\bottomrule
	\end{tabular}
\end{table}

\subsubsection{Postprocessing}

For the postprocessing part, only $a_s$ and $b_s$ are used to
calculate energy. It is noted that $a_s$ and $b_s$ have multiple
channels (multiple scalar values for single atom and single
bond). First, single scalar values for each atom and bond are
calculated by,
\begin{equation}
\begin{split}
    a_{\mathrm{last}}=\mathrm{lin}(a_s), \;\;\;
    b_{\mathrm{last}}=\mathrm{lin}(b_s),
\end{split}
\end{equation}
where the number of channels of $a_{\mathrm{last}}$ and $b_{\mathrm{last}}$ are one.

Then, $a_{\mathrm{last}}$ and $b_{\mathrm{last}}$ are summed along all
atoms and bonds. The obtained single scalar value is the output value
(total energy $E$) of this model.

\begin{equation}
 E=\sum_{\mathrm{atoms}}a_{\mathrm{last}}+\sum_{\mathrm{bonds}}b_{\mathrm{last}}.
\end{equation}

\subsection{Local interaction block: overview}

\begin{figure}[th]
	\centering
	\includegraphics[width=0.9\linewidth,trim=00 00 00 00]{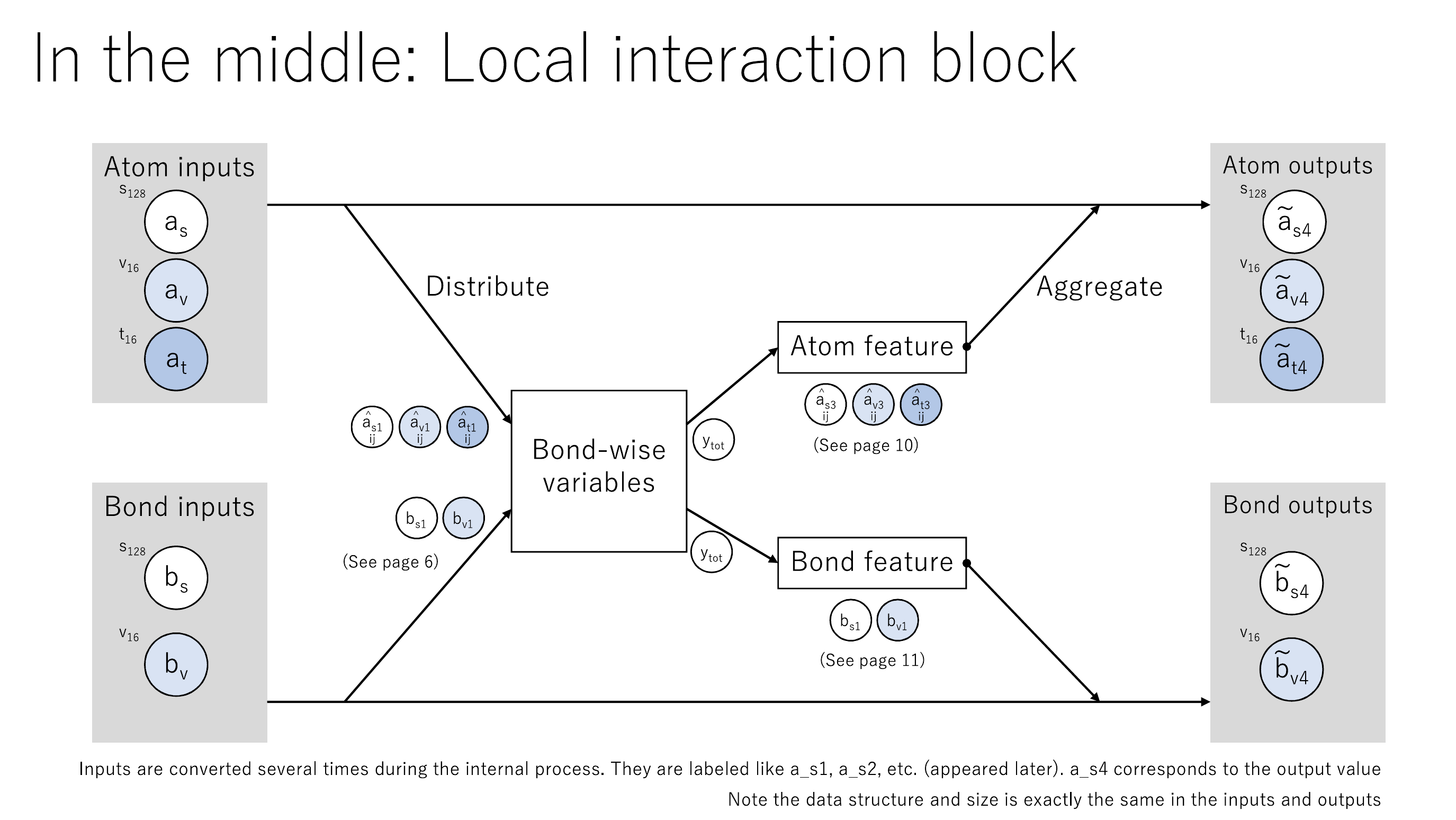}
	\caption{Local interaction block: overview}
	\label{fig:SILocalInteractionBlockOverview}
\end{figure}

This section shows the overview of the calculation flow of local
interaction block. The detail of each calculation block will be
described in the later sections.

First, several operations are applied to the atom-wise inputs ($a_s$,
$a_v$, $a_t$) and the bond-wise inputs ($b_s$, $b_v$). Those newly
created values during the local interaction block are named $a_{s1}$
or $a_{v1}$.

Then, atom-wise values are distributed to the corresponding bonds. It
is noted that there are always two atom-wise values for single
bond. Those distributed values are
concatenated with bond-wise values with keeping required
invariances. The bond shape values ($r_s$, $r_v$) are also used
here. Then, the new bond-wise value named $y_{\rm tot}$ is created
using those values.

After that, new atom-wise variables and bond-wise variables are
created using $y_{\mathrm{tot}}$. Those values are added to the atom
input values and bond input values. Finally, the same shapes of values
as the input values ($a_s$, $a_v$, $a_t$, $b_s$, $b_v$) are returned.

\subsection{Local interaction block: preprocessing}

\begin{figure}[th]
	\centering
	\includegraphics[width=0.9\linewidth,trim=00 00 00
          00]{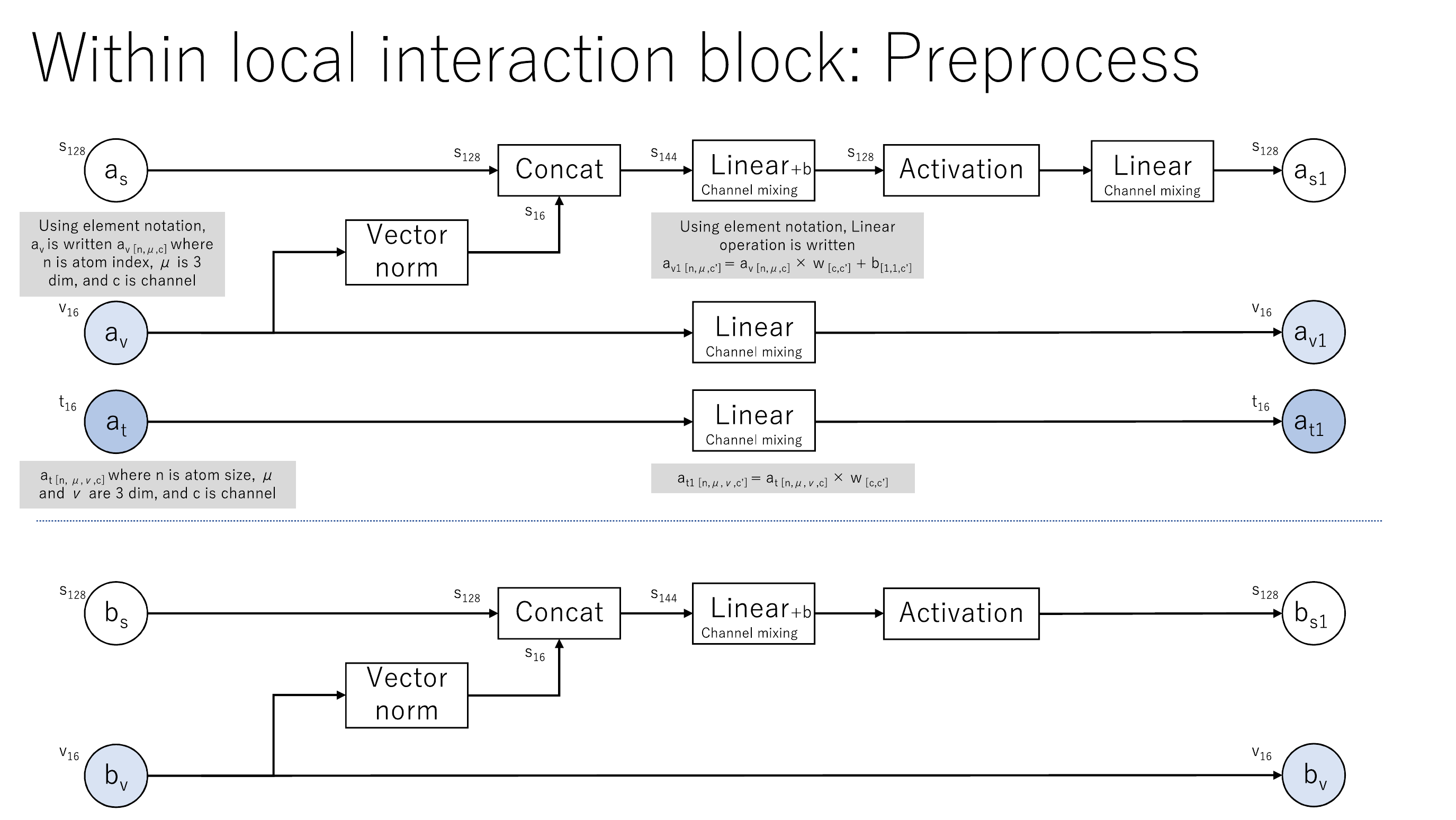}
	\caption{Local interaction block: preprocessing}
	\label{fig:SILocalInteractionBlockPreprocessing}
\end{figure}

As described before, the local interaction block receives $a_s$,
$a_v$, $a_t$, $b_s$, and $b_v$ as inputs. First, several linear and
nonlinear functions are applied for each values.

\begin{equation}
\begin{split}
a_{s1} &= \mathrm{lin}(\mathrm{act}(\mathrm{lin}(\mathrm{con}(a_s, \mathrm{norm}(a_v))))), \\
a_{v1} &= \mathrm{lin}(a_v), \\
a_{t1} &= \mathrm{lin}(a_t), \\
b_{s1} &= \mathrm{act}(\mathrm{lin}(\mathrm{con}(b_s, \mathrm{norm}(b_v)))).
\end{split}
\end{equation}

Here, concat means the values are concatenated along the channel axis.

For linear channel mixing, the linear operation is not applied along
the space dimension axis but along the channel axis.  It is noted that
the raw components of vector and tensor values should not be summed,
multiplied independently, or combined with other scalar values since
the those components depend on the basis vectors of the coordination
system. On the other hand, linear function along channel axis, scalar
multiplication, inner product (including vector norm), and tensor
product are allowed operations.

In this paper, vector norm means the L2-norm of vector along dimension
axis. The result values can be treated as the scalar values.

\subsection{Local interaction block: distribution} \label{sec:distribution}

\begin{figure}[th]
	\centering
	\includegraphics[width=0.9\linewidth,trim=00 00 00
          00]{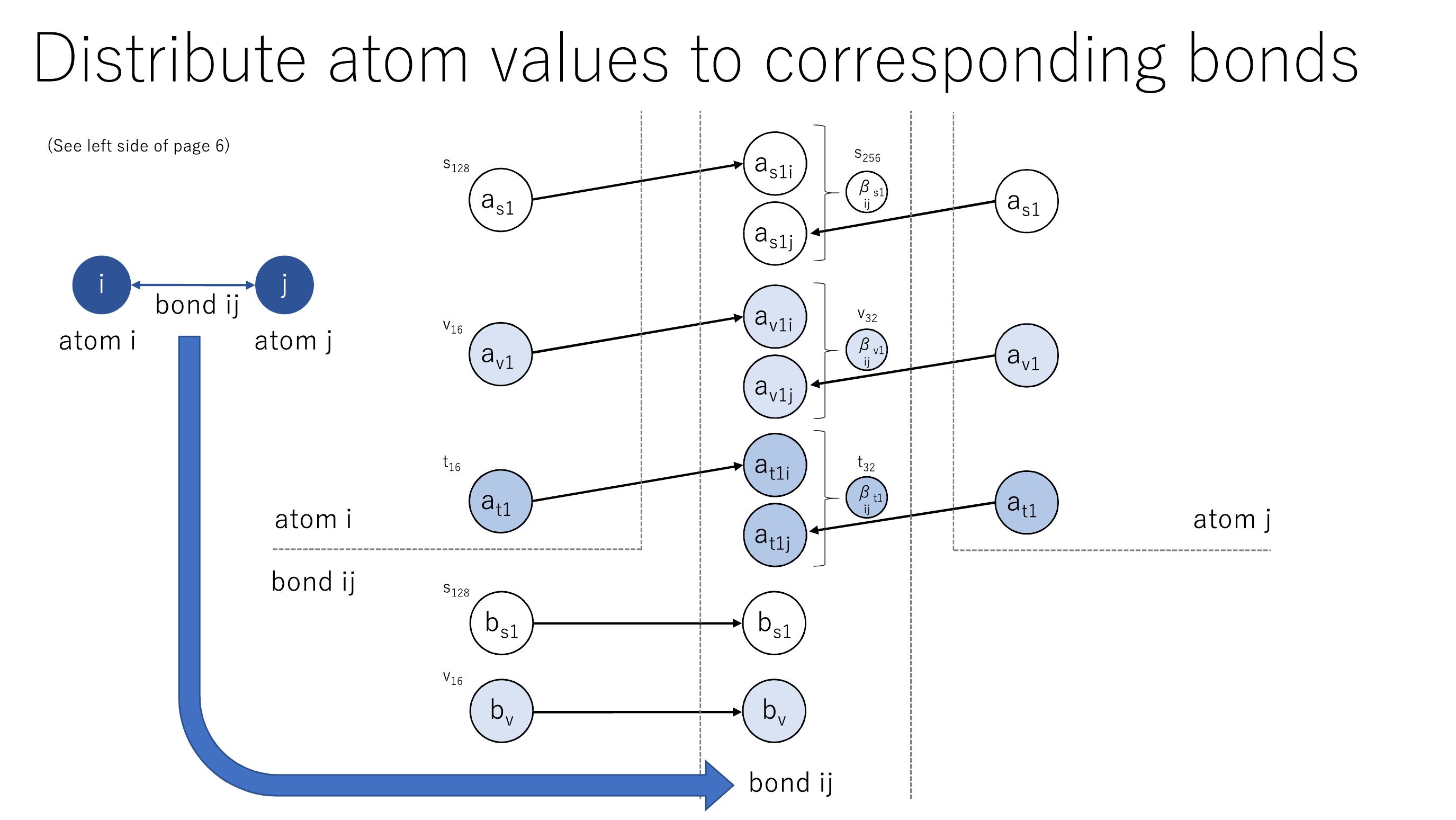}
	\caption{Local interaction block: distribution}
	\label{fig:SILocalInteractionBlockDistribution}
\end{figure}

The atom-wise variables $a_{s1}$, $a_{v1}$, and $a_{t1}$ are
distributed to the corresponding bonds.  It is noted that there are
always two atom-wise values for single bond. We labeled them by $i$
and $j$ as described before.

To clarify that the distributed values corresponds to the bonds, we
name the distributed atom-type values as $\beta_{s1}$, $\beta_{v1}$,
and $\beta_{t1}$. Since there are two corresponding atoms ($i$ and
$j$) for single bond, there are two $\beta$ values such as
$\beta_{s1i}$ and $\beta_{s1j}$. We write
$\beta_{s1\left\{i,j\right\}}$ when the same operations are applied
along $i$ and $j$.

We now have $\beta_{s1\left\{i,j\right\}}$,
$\beta_{v1\left\{i,j\right\}}$, $\beta_{t1\left\{i,j\right\}}$,
$b_{s1}$, and $b_v$. Independently, we have $r_s$ and $r_v$. All of
those values are bond-wise values.

\subsection{Create bond-wise values: preparation}

\begin{figure}[th]
	\centering
	\includegraphics[width=0.9\linewidth,trim=00 00 00
          00]{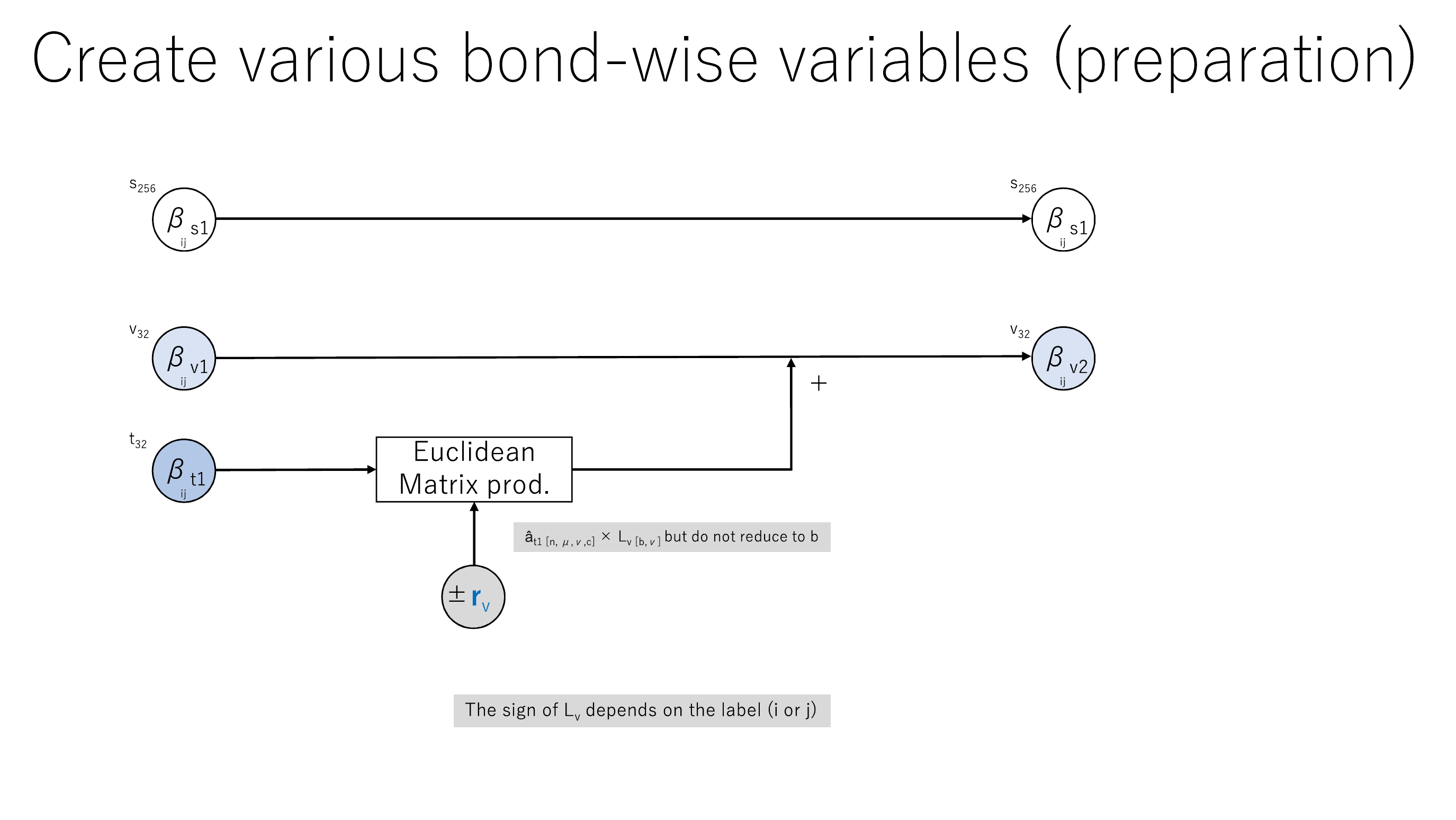}
	\caption{Create bond-wise values: preparation}
	\label{fig:SICreateBondWiseValuesPreparation}
\end{figure}

Tensor value $\beta_{t1\left\{i,j\right\}}$ is squashed into vector
values by taking inner product with $r_v$, and then summed to
$\beta_{v1\left\{i,j\right\}}$.

\begin{equation}
    \beta_{v2\left\{i,j\right\}}=\beta_{v1\left\{i,j\right\}}\pm_{ij}\beta_{t1\left\{i,j\right\}}\cdot r_v.
\end{equation}

It is noted that the sign of $r_v$ depends on the order of the atomic
label ($i$ or $j$). Therefore, to keep the $i$-$j$ order invariance,
the sign should be flipped when the operation is applied to
$j$-related values.  We use the symbol $\pm_{ij}$ for that case. In
the figure, the $i$-$j$ order sensitive values are highlighted as blue
characters and lines.

\subsection{Create bond-wise values: create various intermediate values}

\begin{figure}[th]
	\centering
	\includegraphics[width=0.9\linewidth,trim=00 00 00
          00]{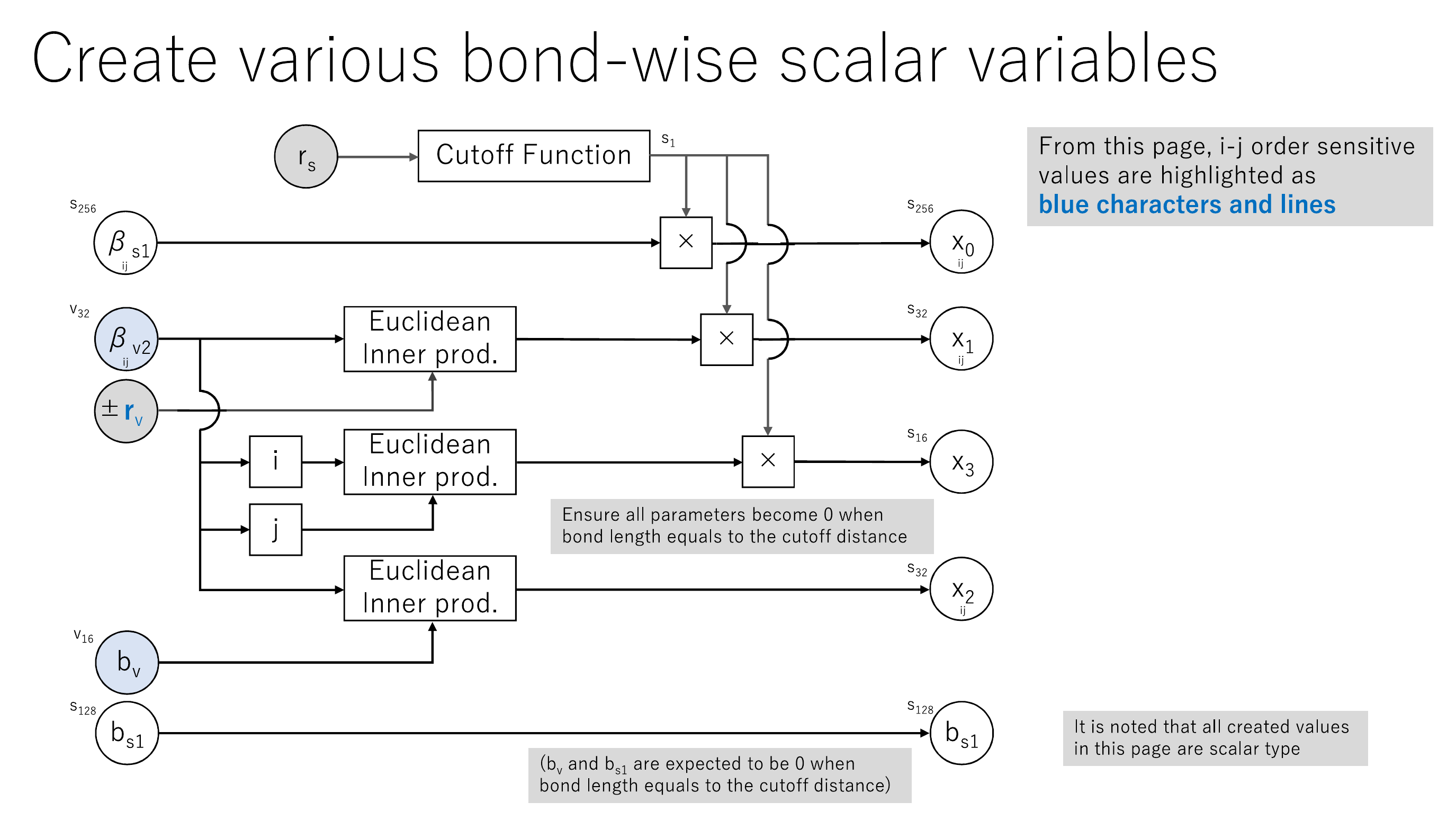}
	\caption{Create bond-wise values: create various intermediate values}
	\label{fig:SICreateBondWiseValuesIntermediate}
\end{figure}

Various bond-type scalar values are calculated by taking the inner
products of vector values.

\begin{equation}
\begin{split}
    x_{0\left\{i,j\right\}} &= \beta_{s\left\{i,j\right\}}\mathrm{cut}(r_s), \\
    x_{1\left\{i,j\right\}} &= \pm_{ij}\beta_{v2\left\{i,j\right\}}\cdot r_v\mathrm{cut}(r_s), \\
    x_{2\left\{i,j\right\}} &= \beta_{v2\left\{i,j\right\}}\cdot b_v, \\
    x_3 &= \beta_{v2i}\cdot\beta_{v2j}\mathrm{cut}(r_s).
\end{split}
\end{equation}

It is noted that $\pm_{ij}$ is used for $r_v$ part again.  The cutoff
function $\mathrm{cut}(r_s)$ is multiplied for
$x_{0\left\{i,j\right\}}$, $x_{1\left\{i,j\right\}}$, and $x_3$ to
ensure that all values are zero when the bond distance equals to the
cutoff distance. It does not be applied to $x_{2\left\{i,j\right\}}$
since bond-related value $b_v$ is assumed to have the same nature.

\subsection{Create bond-wise values: concatenation}

\begin{figure}[th]
	\centering
	\includegraphics[width=0.9\linewidth,trim=00 00 00
          00]{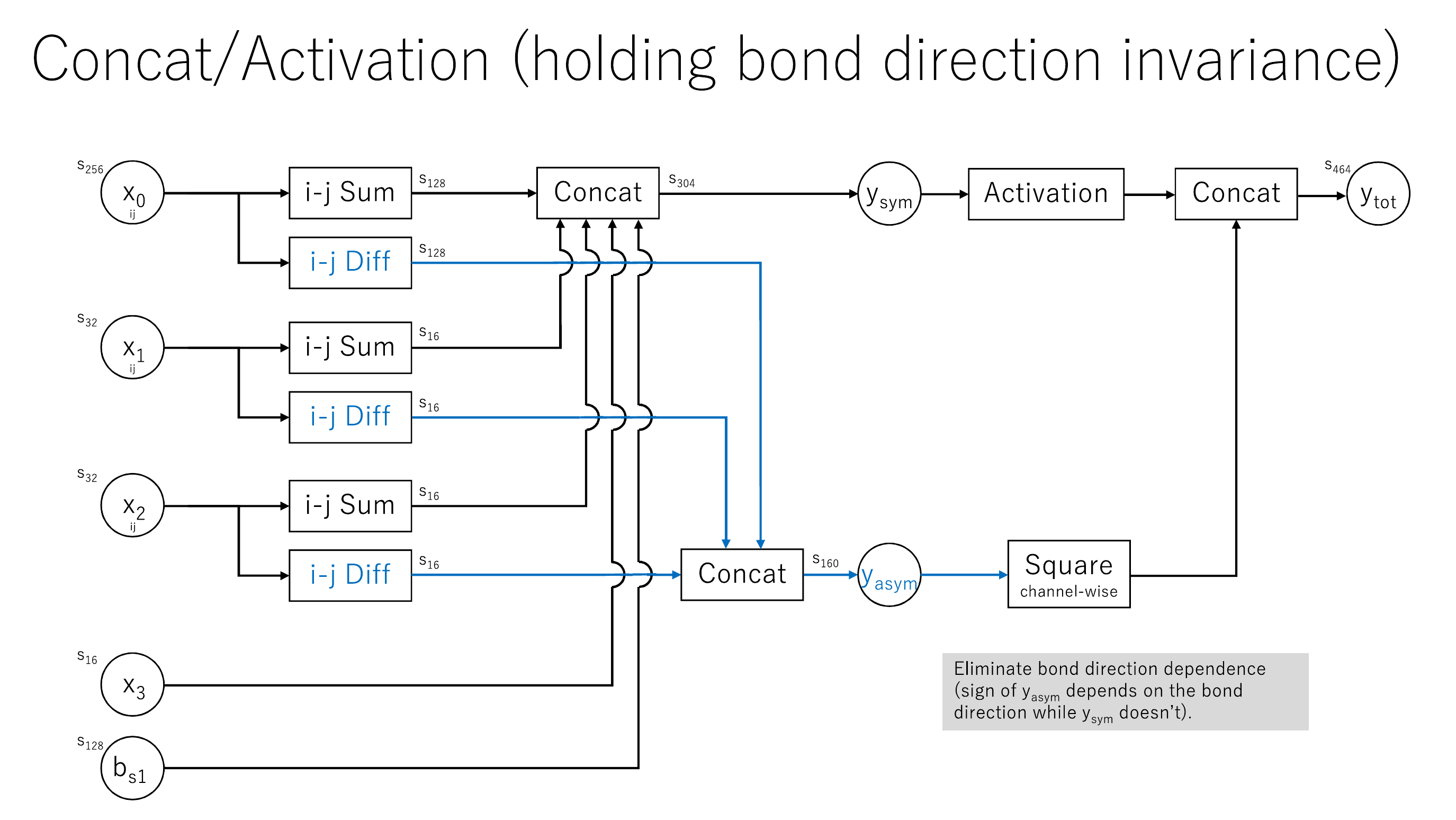}
	\caption{Create bond-wise values: concatenation}
	\label{fig:SICreateBondWiseValuesConcatenation}
\end{figure}

The goal of this section is to create the unified bond-wise value
$y_{\mathrm{tot}}$ from the previously created values. The obtained
scalar values are $x_{0\left\{i,j\right\}}$,
$x_{1\left\{i,j\right\}}$, $x_{2\left\{i,j\right\}}$, $x_3$, and
$b_{s1}$.

The thing left to be done is to eliminate $i$-$j$ order dependence.
It is noted that the values of $x_{0\left\{i,j\right\}}$ swap if we
swap atom $i$ and $j$. In this architecture, we first calculate the
summation and difference ($x_{0i}+x_{0j}$ and $x_{0i}-x_{0j}$). The
former one does not have order dependence and the latter one has order
dependence only on its sign. Therefore, applying the even function for
the latter one removes the order dependence. Here, we use the square
function. The same treatment is carried out for
$x_{1\left\{i,j\right\}}$ and $x_{2\left\{i,j\right\}}$.

\begin{equation}
\begin{split}
y_{\mathrm{sym}} &= \mathrm{lin}(\mathrm{con}(x_{0i}+x_{0j}, x_{1i}+x_{1j}, x_{2i}+x_{2j}, x_{3}, {\bf e}_{s1})), \\
y_{\mathrm{asym}} &= \mathrm{lin}(\mathrm{con}(x_{0i}-x_{0j}, x_{1i}-x_{1j}, x_{2i}-x_{2j})), \\
y_{\mathrm{tot}} &= \mathrm{act}(y_{\mathrm{sym}})+\left(y_{\mathrm{asym}}\right)^{2},
\end{split}
\end{equation}
where $\left(y_{\mathrm{asym}}\right)^{2}$ means element-by-element
square. $y_{\mathrm{tot}}$ is considered to represent the state of the
bond.

In the figure, we highlighted the order-sensitive calculation flow as
blue characters and lines.

\subsection{Local interaction block: create atomic values for update}

\begin{figure}[th]
	\centering
	\includegraphics[width=0.9\linewidth,trim=00 00 00
          00]{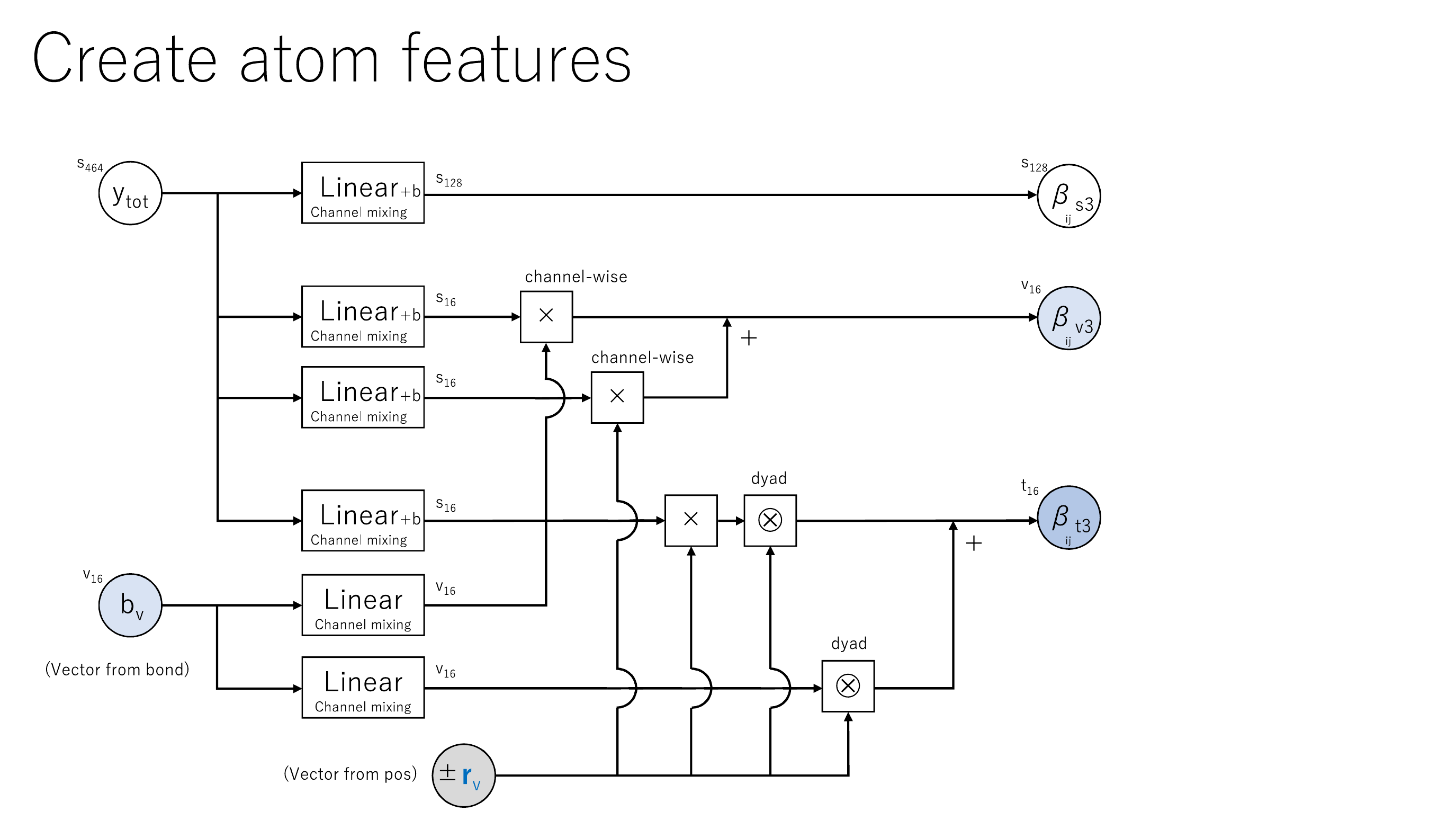}
	\caption{Local interaction block: create atomic values for update}
	\label{fig:SILocalInteractionBlockAtomicValuesUpdate}
\end{figure}

Using $y_{\mathrm{tot}}$, various values which will be accumulated to
atom-wise values and bond-wise values are created. Atom-type variables
are calculated by,

\begin{equation}
\begin{split}
    \beta_{s3\left\{i,j\right\}} &= \mathrm{lin}(y_{\mathrm{tot}}), \\
    \beta_{v3\left\{i,j\right\}} &= \mathrm{lin}(y_{\mathrm{tot}})\mathrm{lin}(b_v)\pm_{ij}\mathrm{lin}(y_{\mathrm{tot}})r_v, \\
    \beta_{t3\left\{i,j\right\}} &= \mathrm{lin}(y_{\mathrm{tot}})r_v\otimes r_v\pm_{ij}\mathrm{lin}(b_v)\otimes r_v.
\end{split}
\end{equation}

\subsection{Local interaction block: create bond values for update}

\begin{figure}[th]
	\centering
	\includegraphics[width=0.9\linewidth,trim=00 00 00
          00]{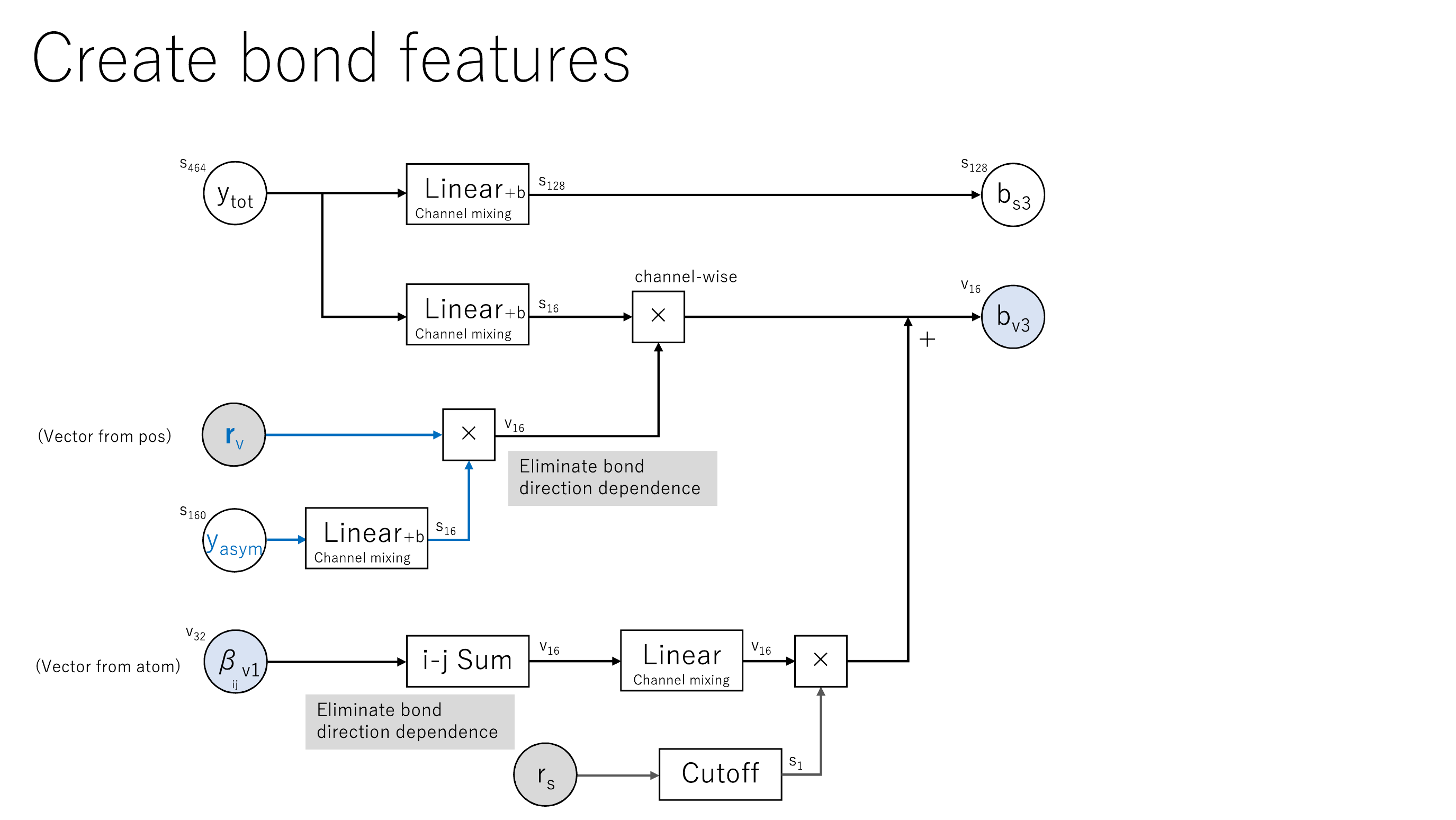}
	\caption{Local interaction block: create bond values for update}
	\label{fig:SILocalInteractionBlockBondValuesUpdate}
\end{figure}

In the same manner to the atom-wise values, bond-wise values are
calculated using ${\bf y}_{\mathrm{tot}}$.

\begin{equation}
\begin{split}
    b_{s3} &= \mathrm{lin}(y_{\mathrm{tot}}), \\
    b_{v3} &= \mathrm{lin}(y_{\mathrm{tot}})\mathrm{lin}(y_{\mathrm{asym}}) r_v+\mathrm{lin}(\beta_{v2i}+\beta_{v2j})\mathrm{cut}(r_s).
\end{split}
\end{equation}

For creating bond vector value $b_{v3}$, $y_{\mathrm{asym}}$ is
introduced to eliminate the $i$-$j$ order dependence.

\subsection{Local interaction block: aggregation}

\begin{figure}[th]
	\centering
	\includegraphics[width=0.9\linewidth,trim=00 00 00
          00]{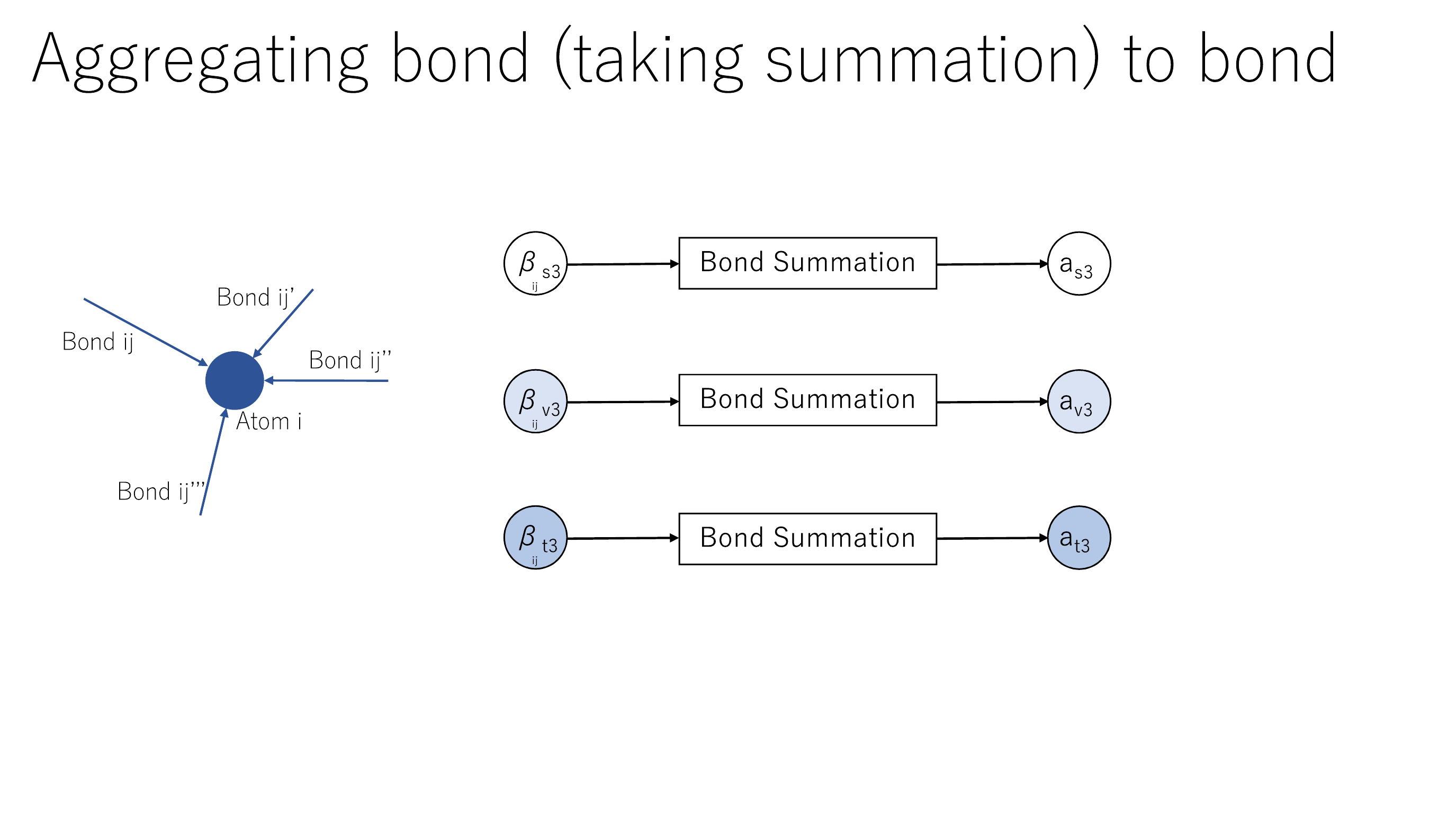}
	\caption{Local interaction block: aggregation}
	\label{fig:SILocalInteractionBlockAggregation}
\end{figure}

$\beta_{s3\left\{i,j\right\}}$, $\beta_{v3\left\{i,j\right\}}$,
$\beta_{t3\left\{i,j\right\}}$ are intended to update the atom-wise
values. However, those values are still bond-wise values and needed to
be aggregated to the corresponding atoms.  This is done by taking the
summation of neighboring bond-wise values to the atoms. This is the
inverse calculation flow to the distribution described in the section
\ref{sec:distribution}

We name the summed atomic values as $a_{s3}$, $a_{v3}$, and $a_{t3}$.

\subsection{Local interaction block: create output values}

\begin{figure}[th]
	\centering
	\includegraphics[width=0.9\linewidth,trim=00 00 00
          00]{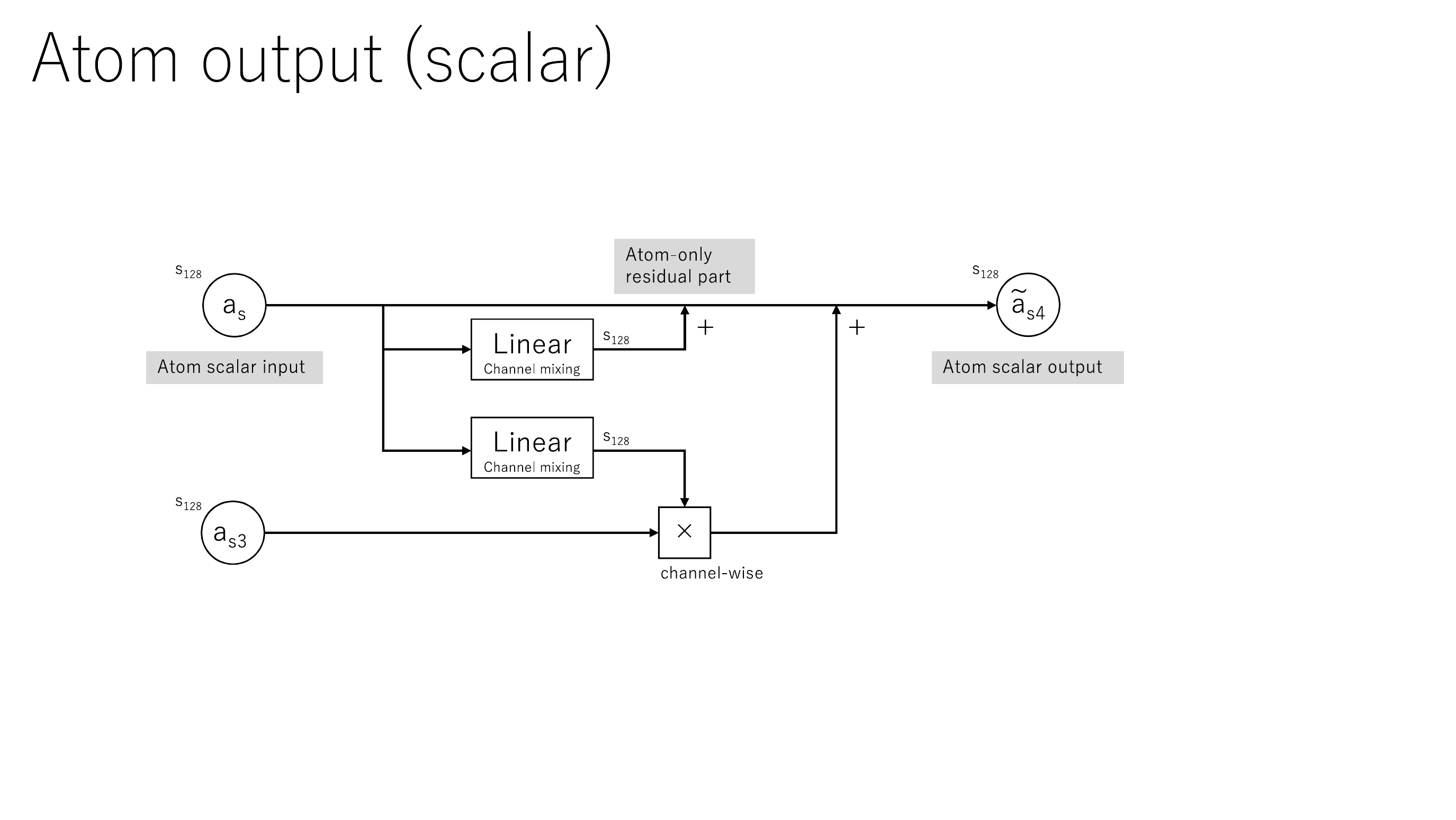}
	\caption{Local interaction block: output 1}
	\label{fig:SILocalInteractionBlockOutput1}
\end{figure}
\begin{figure}[th]
	\centering
	\includegraphics[width=0.9\linewidth,trim=00 00 00
          00]{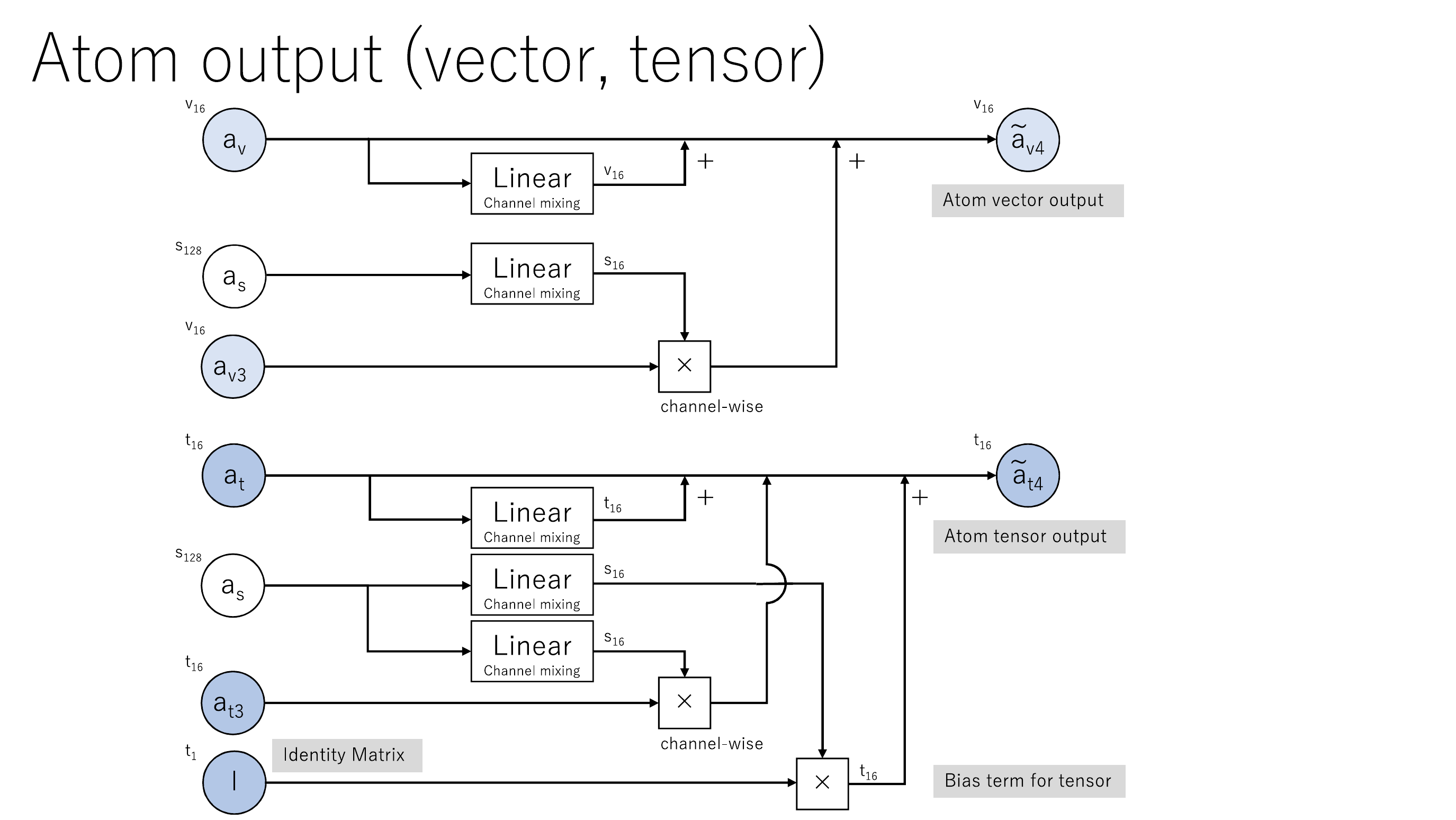}
	\caption{Local interaction block: output 2}
	\label{fig:SILocalInteractionBlockOutput2}
\end{figure}
\begin{figure}[th]
	\centering
	\includegraphics[width=0.9\linewidth,trim=00 00 00
          00]{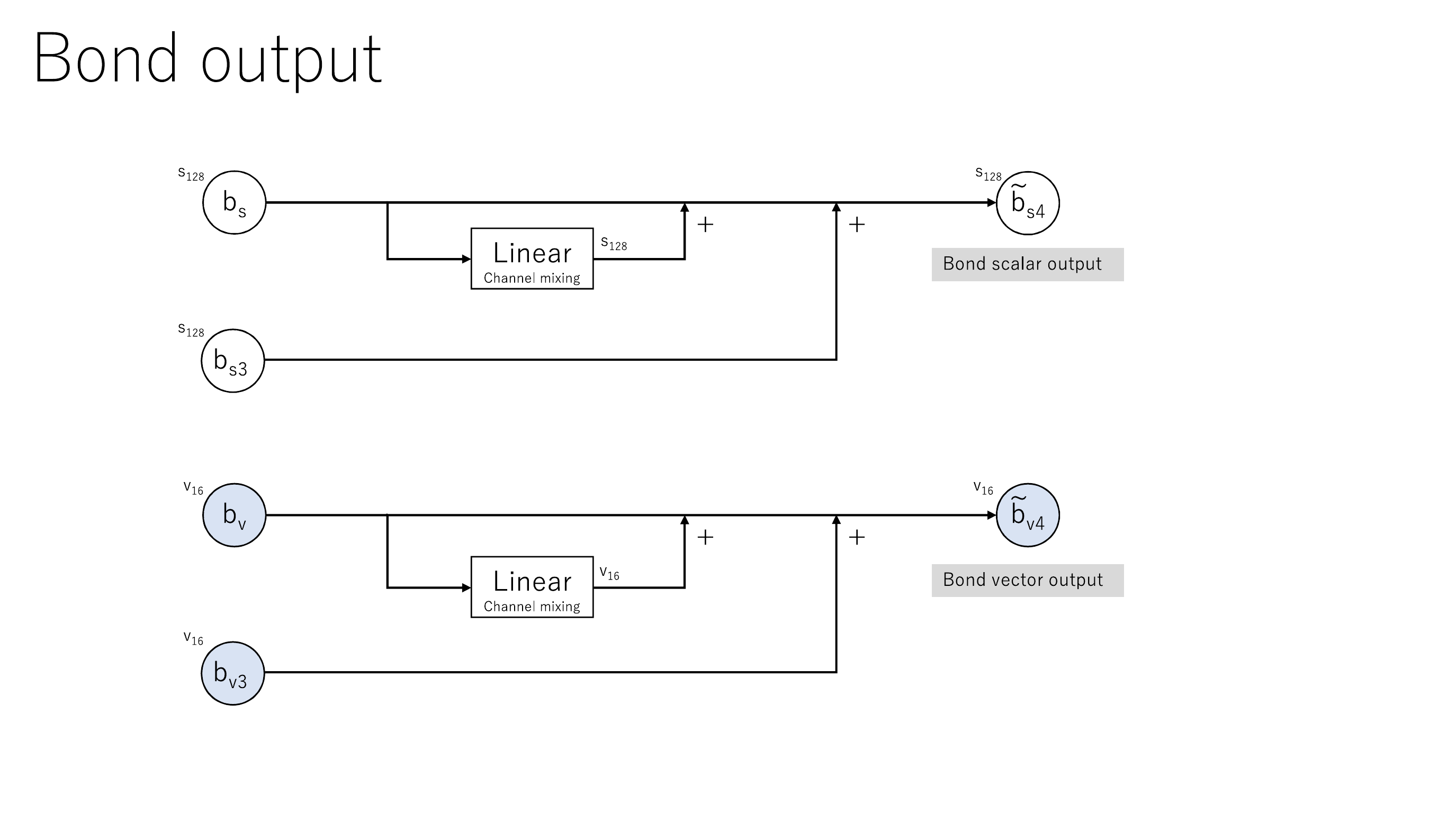}
	\caption{Local interaction block: output 3}
	\label{fig:SILocalInteractionBlockOutput3}
\end{figure}

Finally, node and edge variables are updated by ResNet-style bypass function.

\begin{equation}
\begin{split}
\tilde{a}_{s4} &= a_s+\mathrm{lin}(a_s)+\mathrm{lin}(a_s)a_{s3}, \\
\tilde{a}_{v4} &= a_v+\mathrm{lin}(a_v)+\mathrm{lin}(a_s)a_{v3}, \\
\tilde{a}_{t4} &= a_t+\mathrm{lin}(a_t)+\mathrm{lin}(a_s)a_{t3}+\mathrm{lin}(a_s)\mathrm{I}, \\
\tilde{b}_{s4} &= b_s+\mathrm{lin}(b_s)+b_{s3}, \\
\tilde{b}_{v4} &= b_v+\mathrm{lin}(b_v)+b_{v3},
\end{split}
\end{equation}

where $\mathbf{I}$ is the identity tensor which is used as the bias
term. The first term is a residual part and the second term is the
structure-independent value update part.

It is noted that $\mathrm{lin}(a_s)$ is multiplied to atom-wise update
part. It is considered to work as a node convolution gate
function. These variables are the final output of the interaction
block and used as the input variables of the next block.

Those five values ($\tilde{a}_{s4}$, $\tilde{a}_{v4}$,
$\tilde{a}_{t4}$, $\tilde{b}_{s4}$, and $\tilde{b}_{v4}$) are the
output values of the local interaction block. They are used for the
input values of the next local interaction block or the postprocess
layer.

\section{TeaNet Philosophy and Training}

With the detailed network laid out in section \ref{sec:architecture},
we now zoom out and discuss the underlying philosophy of TeaNet. We
would like to show the correspondence between existing physics-based
potentials (EAM and Tersoff-type angular-dependent potentials) and
GCN, in section \ref{sec:eam} and \ref{sec:tensor}, respectively. We
show that the Tersoff-type angular-dependent bond-order potential can
also be rewritten as the graph convolution by incorporating the
Euclidean tensor variables into GCN architecture. This means that the
rank-2 tensors empower GCN to treat the spatial information naturally
while keeping frame-rotation, reflection, and translation
equivariances. We also show the necessity of tensor values for
transferring spatial information in graph convolution
architecture. Then in section \ref{sec:minimize}, we introduce the
constraint which enables the model to be stacked deeper and to improve
the accuracy. Then we explain the analogy of this constraint with the
energy relaxation procedure of the charge-transfer-type IP, which is
known as charge equilibration (QEq)
method\cite{doi:10.1021/j100161a070}

\subsection{Rewriting EAM potential as graph convolution} \label{sec:eam}

\begin{figure}[th]
	\centering
	\includegraphics[width=0.9\linewidth,trim=00 00 00 00]{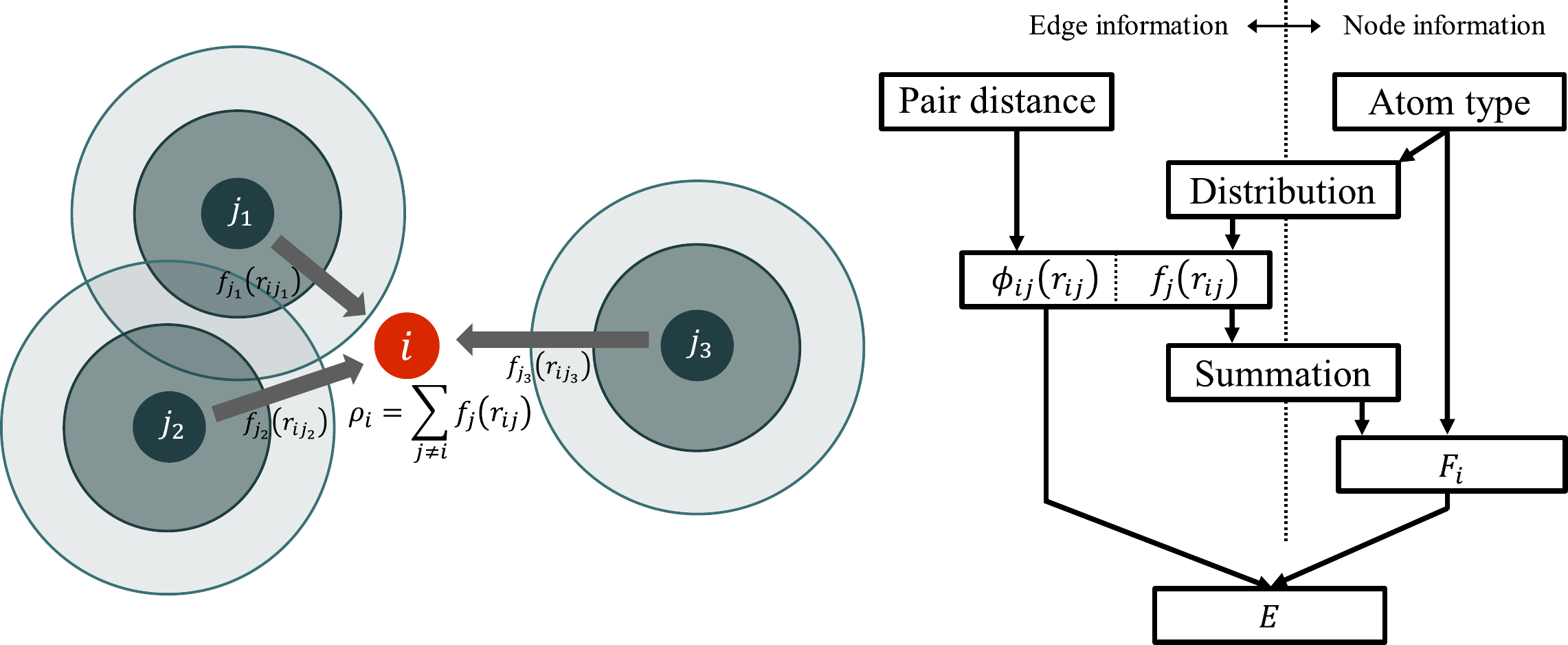}
	\caption{EAM potential represented as a graph convolution. Left: Schematic of the summation operation. Right: Corresponding network model. }
	\label{fig:eam}
\end{figure}

The EAM potential \cite{PhysRevB.29.6443} incorporates the
concept of electron density in a shallow 1-layer network. In EAM, the total energy, $E$, is calculated as:
\begin{equation}
\begin{split}
  E&=\frac{1}{2}\sum_{i}\sum_{j\neq i}\phi_{ij}\left(r_{ij}\right)+\sum_i F_{i}\left(\rho_i\right), 
 \label{eq:eam} \\
  \rho_i&=\sum_{j\neq i}f_j\left(r_{ij}\right),
\end{split}
\end{equation}
where $i, j$ are the atom labels and $\phi_{ij}$, $F_{i}$, $f_{j}$,
and $r_{ij}$ are functions describing the two-body energy, the
embedding energy, the electron charge, and the interatomic distance,
respectively. In EAM potential, $\rho_i$ which corresponds to the
background electron density at atom $i$ is calculated by the summation
of pairwise function. It can be expressed as a single-layer graph
convolution (see Fig. \ref{fig:eam}).

The EAM potential can be translated as a shallow GCN as follows: The
atomic information (on the nodes) is distributed to the corresponding
bonds $r_{ij}$. Then, the bond-wise values
($\phi_{ij}\left(r_{ij}\right), f_i\left(r_{ij}\right)$) are
calculated. A part of them ($f_i\left(r_{ij}\right)$) are summed to
the corresponding atoms and atom-site nonlinear function
($F_{i}\left(\rho\right)$) is applied. It is noted that EAM potential
has the required invariances such as permutation, pair order, and
isometry.

The calculation flow of our graph convolution layer follows the idea
of the EAM potential. First, the atom-wise values are
calculated. Then, they are distributed into the corresponding bonds
and the bond-wise values are calculated by combining atom-wise values
and bond-wise values.  After that, the calculated bond-wise values are
transferred into the corresponding atoms and update the atom-wise
values.

To accumulate the edge information ($f_j\left(r_{ij}\right)$) into
nodes, the embedding function ($F_{i}$) plays an important role. In
EAM potential, $F_{i}$ represents the interaction between certain
atoms and the surrounding electron density. Therefore, from a physics
standpoint, the embedding function is essential in the network
architecture. We call this the ``node gate'' function. The effect of
the node gate function on prediction accuracy is presented in section
\ref{sec:train_result}.

\subsection{Translating bond angle interaction into graph convolution and embedding vector and tensor values} \label{sec:tensor}

Generally speaking, atomic interactions depend on the bond angle
between interacting atoms. For example, H${}_2$O and NH${}_3$
molecules are stabilized at a certain bond angle. Diamond comprises a
tetrahedral network. These angular dependencies are generated by the
interaction between electron orbitals \cite{QianLQWCYHY08}.

When embedding spatial information in the network architecture,
satisfying invariance requirements can be challenging. The energy
should be invariant to the rotation of the basis vectors.  Invariance
is handled differently in different models. One solution is to limit
the input data to only the bond length.  SchNet
\cite{schutt2017schnet} and PhysNet
\cite{doi:10.1021/acs.jctc.9b00181} uses bond length only. Deep tensor
neural networks (DTNN) \cite{schutt2017quantum} and deep potential
molecular dynamics (DPMD) \cite{PhysRevLett.120.143001} also maintain
the rotational invariance by using bond length. However, bond-bond
interactions usually depends on the higher-order geometric information
such as bond angle or dihedral angle, and its relation to the bond
length can be weak. The detailed discusssion is in Appendix
\ref{sec:app:length_weak}.  Since the solution of using the raw values
of vector components as the input values loses the rotation
invariance, it is not appropriate for a molecular dynamics simulation.

Many existing IPs involve bond angles directly. For example, the
Stillinger-Weber potential \cite{PhysRevB.31.5262} has a three-body
energy function. Bond-order-type potentials, such as the Tersoff
potential \cite{PhysRevB.37.6991, PhysRevB.39.5566}, possess a
bond-order term consisting of the three-body angular-dependent
term. Some machine learning-based models give similar solutions. The
Behler-Parrinello neural network (BPNN) \cite{PhysRevLett.98.146401}
calculates the three-body symmetry functions.

However, the bond angles correspond to neither nodes nor edges but
rather to three-body atom combinations. Therefore, they should be
combined and converted into representative node and edge values during
the convolution operation, which requires the use of ad-hoc functions
such as symmetry functions.  Another problem is the lack of long-range
interaction in the three-body angle term. In GCN, local information
can transfer to farther nodes through the convolutional
layers. Transfer of the angle information is also desirable. For
example, the directional electronic orbitals of the $\pi$ bonds can be
extensively spread. However, convolution of the angular information at
the node crushes the angle information and prevents its propagation.

Here, we show that the angle-dependent three-body convolution
algorithm can be naturally expressed as a normal node-and-edge
convolution operation using Euclidean vector and second-order tensor
values. This means that the model can have local spatial information
and propagate it to farther nodes, and to interact with them at nodes
while keeping rotational invariances. This is achieved by rewriting
the Tersoff-type angle-dependent bond-order function as a convolution
operation.
 
The Tersoff-type angle-dependent term $\zeta_{ij}$ can be written as
\begin{equation}
\begin{split}
E&=\frac{1}{2}\sum_{i,j\neq i}\phi_{\rm A}\left(r_{ij}\right)+\frac{1}{2}\sum_{i,j\neq i}b\left(\zeta_{ij}\right)\phi_{\rm B}\left(r_{ij}\right), \\
\zeta_{ij}&=\sum_{k\neq i,j} G\left(\theta_{ijk}\right) H\left(r_{ij}, r_{ik}\right),
\end{split}
\end{equation}
where $i$, $j$, and $k$ are the atom labels; $\theta_{ijk}$ is the angle between bonds $ij$ and $ik$; $r_{ij}$ and $r_{ik}$ are the bond lengths, and $\phi_{\rm A}$, $\phi_{\rm B}$, $b$, $G$, and $H$ are various functions.
In some Tersoff-type potentials \cite{doi:10.1063/1.4965863, PhysRevB.97.125411}, the $\zeta_{ij}$ term is expressed as
\begin{equation}
\zeta_{ij}=\sum_{k\neq i,j} \left[c+d\left\{h-\cos\left(\theta_{ijk}\right)\right\}^2\right]f_c\left(r_{ik}\right)\exp\left[\lambda \left(r_{ij}-r_{ik}\right)\right], \label{eq:tersoff2}
\end{equation}
where $f_{\rm c}$ is the cutoff function and $c$, $d$, $h$, and $\lambda$ are the parameters. After expanding the factors and converting the parameters, Eq. (\ref{eq:tersoff2}) is transformed to
\begin{equation}
\begin{split}
\zeta_{ij}&=\exp\left(\lambda r_{ij}\right)\sum_{k\neq i,j} \left[g_0+g_1\cos\left(\theta_{ijk}\right)+g_2\cos^2\left(\theta_{ijk}\right)\right]f_{\rm c}\left(r_{ik}\right)\exp\left(-\lambda r_{ik}\right) \\
&=\exp\left(\lambda r_{ij}\right)\sum_{k\neq i,j} \left[g_0+g_1\hat{\bf r}_{ij}\cdot\hat{\bf r}_{ik}+g_2\left(\hat{\bf r}_{ij}\cdot\hat{\bf r}_{ik}\right)^2\right]f_{\rm c}\left(r_{ik}\right)\exp\left(-\lambda r_{ik}\right) \\
&=\exp\left(\lambda r_{ij}\right)\sum_{k\neq i,j} \left[g_0+g_1\hat{\bf r}_{ij}\cdot\hat{\bf r}_{ik}+g_2\left(\hat{\bf r}_{ij}\otimes\hat{\bf r}_{ij}\right):\left(\hat{\bf r}_{ik}\otimes\hat{\bf r}_{ik}\right)\right]f_{\rm c}\left(r_{ik}\right)\exp\left(-\lambda r_{ik}\right) \\
&=\exp\left(\lambda r_{ij}\right)\sum_{k\neq i} \left[g_0+g_1\hat{\bf r}_{ij}\cdot\hat{\bf r}_{ik}+g_2\left(\hat{\bf r}_{ij}\otimes\hat{\bf r}_{ij}\right):\left(\hat{\bf r}_{ik}\otimes\hat{\bf r}_{ik}\right)\right]f_{\rm c}\left(r_{ik}\right)\exp\left(-\lambda r_{ik}\right) \\
&\ \ \ \ -\left(g_0+g_1+g_2\right)f_{\rm c}\left(r_{ij}\right) \\
&=g_0\exp\left(\lambda r_{ij}\right)\left[\sum_{k\neq i} f_{\rm c}\left(r_{ik}\right)\exp\left(-\lambda r_{ik}\right)\right] \\
&\ \ \ \ +g_1\exp\left(\lambda r_{ij}\right)\hat{\bf r}_{ij}\cdot\left[\sum_{k\neq i} \hat{\bf r}_{ik}f_{\rm c}\left(r_{ik}\right)\exp\left(-\lambda r_{ik}\right)\right] \\
&\ \ \ \ +g_2\exp\left(\lambda r_{ij}\right)\left(\hat{\bf r}_{ij}\otimes\hat{\bf r}_{ij}\right):\left[\sum_{k\neq i} \left(\hat{\bf r}_{ik}\otimes\hat{\bf r}_{ik}\right)f_{\rm c}\left(r_{ik}\right)\exp\left(-\lambda r_{ik}\right)\right] \\
&\ \ \ \ -\left(g_0+g_1+g_2\right)f_{\rm c}\left(r_{ij}\right),
\end{split}
\end{equation}
where $\hat{\bf r}_{ij}$ and $\hat{\bf r}_{ik}$ are the unit
vectors. The symbols ``$\cdot$,'' ``$:$,'' and ``$\otimes$'' denote
the inner product, the Frobenius inner product, and the tensor product
(dyad) of two vectors, respectively. Since all summation terms are
written without $j$, they can be calculated by the convolution
operation. As a result, the Tersoff-type potential function can be
written as a two-layered neural network.  The necessity of the Rank-2
tensors for the angle interaction using convolution operation and its
physical meaning and comparison with spherical harmonics-based methods
are shown in the Appendix \ref{sec:app:tensor}.

Based on this discussion, we introduce both vectors and tensors into
the network. Each node array contains scalar, vector, and tensor
values, whereas each edge array contains scalar and vector values. A
relative position vector is also provided as an input value. The
effects of tensor values on prediction accuracy are presented in the
section \ref{sec:train_result}.  See section \ref{sec:specification}
for the details of the implementation.

\subsection{Improvement of stacked GCN accuracy inspired of iterative
  energy minimization process}
\label{sec:minimize}

Like in existing GCNs, the local interaction block can be stacked
multiple times. However, as frequently seen in NN training, we
observed the increase of the number of layers always brings the
instability during the learning procedure. Therefore, it was hard to
improve the accuracy by increasing the number of layers of our model
in practice.

Here, we found a method to reduce this instability significantly. The
key idea is to initialize and to make a constraint that all middle
layers have the same NN parameters at the initial stage of
training. One can find similarities to the recurrent GCN architecture
\cite{4700287}. Another essential point is to apply the residual
network (ResNet) architecture. Interestingly, we found that the
accuracy was improved by increasing the number of layers up to
16. (see architectural details in section \ref{sec:train_result}).  It
is said that improving the expression power of GCN is hard by
increasing the depth size \cite{DBLP:conf/iclr/PapernotAEGT17}. This
was also true in atomistic system in practice. Many GCN models in
atomistic system also have up to 6 convolution layers
\cite{schutt2017schnet, doi:10.1021/acs.jctc.9b00181,
  PhysRevLett.120.143001}. In addition, making the constraint to set
the all middle layers have the same parameters can be thought to
enforce them to behave the identical nonlinear transformation, which
seems to reduce the expressive ability of the entire
network. Therefore, it is not strange to think that this method does
not contribute to the accuracy. It is noted that this constraint was
came from an analogy with physics. In this section, we explain the
analogy and introduce the insight why the deeply stacked model can
improve the accuracy even it is GCN.

Limiting the number of GCN layers to one means the node's information
can be determined only by the neighboring nodes. In the analogy with
GCN and EAM potential, this corresponds to the assumption that the
electron state (density) of the atom can be calculated only by
surrounding atoms.
Although the assumption works well for certain systems, it is not
physically correct picture in general, as seen by the long-ranged
nature of the dielectric response function in DFT\cite{HeV01}. The
charge transfer effect plays important roles in chemical reactions.
The actual electron states are determined so that the energy of the
entire system is minimized. DFT calculates the ground state of the
electron density by an iterative procedure. To incorporate such
long-ranged propagation of information, charge-transfer-type IPs
\cite{doi:10.1021/j100161a070, doi:10.1021/jp004368u,
  PhysRevB.75.085311, doi:10.1063/1.4965863}, which model the
deviation of the electron density and minimize the energy of the
system with respect to the charge distribution, are being actively
developed.

In charge-transfer-type IPs, the energy minimization involves implicit
matrix-vector equations solved by matrix inverse calculation
\cite{doi:10.1021/j100161a070, doi:10.1021/jp004368u} or solved by
repeatedly updating the charge distribution using the gradient-based
method \cite{PhysRevB.75.085311}. If the number of iterations is
fixed, this iterative procedure could be written as a feed-forward
data flow model.  It is noted that iterative total energy minimization
reproduces the physically reasonable long-range interactions. A
well-known example is the Green's function solution that can be
represented by a matrix-vector equation ${\bf A}{\bf x}={\bf b}$: even
though ${\bf A}$ is a sparse matrix (local interactions), the inverse
${\bf A}^{-1}$ is dense and resembles long-range
interactions. However, by iteratively solving ${\bf A}{\bf x}={\bf b}$
with Krylov subspace method $\{{\bf b}, {\bf Ab}, {\bf A}^2{\bf b},
{\bf A}^3{\bf b}, ..., {\bf A}^n{\bf b}\}$, one can achieve excellent
approximant to the long-range interaction, which is akin to an
$n$-layer neural network with identical weights.

It should be noted the importance of the residual network architecture
in the above discussion. The residual network (ResNet)
\cite{he2016deep} \deleted{and its derivatives} have recently emerged
in the fields of image recognition, as have other machine-learning
tasks, including object detection \cite{he2017mask}, machine
translation \cite{wu2016google}, and speech synthesis
\cite{van2016wavenet}. ResNet’s core idea is to ``bypass'' the output
values from the middle layers and add them directly to the lower layer
to avoid gradient disappearance during back
propagation. Interestingly, previous studies interpreted the ResNet
architecture as an explicit Euler method of ordinary and partial
differential equations \cite{lu2017beyond, AAAI1816517,
  chen2018neural}. In this section, we associated the stack of the
local interaction blocks using residual network connection with
charge-transfer energy minimization calculation of IPs.

\subsection{Data collection}
\label{sec:data_collection}

Since our target to develop an universal IP with applicability to
arbitrary structures, the dataset is required to cover the wide range
of phase space as much as possible. One solution is to increase the
number of data points. Another requirement is to secure the diversity
of data points.

The dataset is created as follows. First, the simulation box is filled
with tens of atoms. The element type is randomly selected from the
first three rows of the periodic table (from H to Ar). The number of
element types and their ratio in one sample is also widely
distributed. The system is heated to high temperature (e.g. 10,000 K),
melted for approximately $100$ femtoseconds, cooled to a setting
temperature, then further annealed for another $100$ femtoseconds by
classical MD to obtain a snapshot. The timestep is 1 femtosecond. This
process is repeated for various temperatures (up to 5,000 K) and
volumes. Then, the reference energy and atomic forces are obtained by
DFT calculations of the snapshots. We consider that this dataset
consists of highly disordered structures, including many types of
local atomic configurations, and thus presents a challenging
task. Furthermore, most of the configurations are far from stable.

In addition, to include realistic structure, we create another dataset
by heating the structures of the molecular dataset of the Materials
Project repository \cite{doi:10.1063/1.4812323} up to 3,000 K. In this
work, we merged those two datasets. The entire dataset contains
approximately 294,000 structures. The size of the dataset at the
double backpropagation process (the corresponding atomic forces of the
294,000 structures) is approximately 7,375,000. Two-hundred randomly
selected structures (including $4962\times3$ atomic force data) are
used for the test dataset exclusively.

We used VASP for DFT calculation. To increase the number of data
points, the relatively fast settings were used. GGA-PBE was used for
the exchange-correlation energy. The Gaussian smearing was used. Spin
polarization is considered. The smearing width $\sigma$ was 0.2
eV. The PREC setting in VASP (used to determine energy cutoff) was set
to Medium. One k-point was applied. We used the same settings among
structures to ensure the energy surface is consistent. Further
expansion of the dataset (e.g. increasing the number of elements,
increasing the number of structures, improvement of the computational
accuracy) is a future task.

The details of dataset is shown in Appendx \ref{sec:app:dataset_detail}.

\subsection{Training procedure} \label{sec:method_training}

The NN hyperparameters are set as follows. The length of the scalar
node and edge arrays is set to be 128. The length of the vector node,
rank-2 tensor node, and vector edge arrays is each set to 16. The
cutoff distance is set to 6 \mbox{\normalfont\AA}. The minibatch size
is 100.

The network is trained by optimizing the combined absolute loss
function (energies and atomic forces) using the Adam optimizer
\cite{kingma2014adam}. As the number of layers increased, frequent
fluctuations were observed in the training error. This instability may
be explained, at least in part, by the roughness of the DFT-calculated
potential energy surface, which is the ground truth of this
task. Small atomic displacements, such as the approaching of two
neighboring atoms, can potentially cause abrupt energy increases.

To resolve this problem, we constrained the parameters of all
intermediate layers in the network to the same values at the initial
stage of the training, as described in section \ref{sec:minimize}.

Finally, the models were trained by stochastic gradient descent with a
small learning rate ($0.1$). The numbers of iterations were set to
450,000 (initial), 450,000 (main), and 20,000 (final) in all cases.

\section{Training results}
\label{sec:train_result}

\subsection{Dependence of accuracy on the number of NN layers}
There are several datasets for atomic systems, without reaction
barrier information. For example, QM7 (GDB7-12)
\cite{PhysRevLett.108.058301} and QM9 (GDB9-14)
\cite{ramakrishnan2014quantum} are composed of equilibrium molecular
data.  In contrast, to reproduce the wide range of energy surface, the
model should reproduce a wide range of structures. Therefore,
evaluations of highly disordered structures including dangling bonds,
overcoordinated atoms, and various disordered bond lengths are
required.  Therefore, we prepare our own dataset of highly disordered
structures using molecular dynamics simulations.  The dataset consists
of the first three rows of the periodic table (from H to Ar).  The
details of the data preparation are shown in section
\ref{sec:data_collection}.

We trained networks of different depths (2, 4, 8, and 16 layers). The
hyperparameters and other settings for training are shown in section
\ref{sec:method_training}.  The results are depicted in Table
\ref{tab:layer_result}.  Increasing the number of layers improved the
network accuracy. No overfitting was observed in any system.  In the
best-performing network (with 16 layers), the mean absolute error
(MAE) of the energy was 19.3 meV/atom.  Our proposed method enables to
construct deeper model which has higher accuracy in the field of GCN.

\begin{table}[th]
	\caption{Regression accuracy of trained networks with various numbers of layers.}
	\label{tab:layer_result}
	\centering
	\begin{tabular}{ccccc}
		\toprule
		\multicolumn{1}{c}{\# layers} & \multicolumn{1}{c}{\# params} & \multicolumn{1}{c}{\shortstack{Test loss \\function [unitless]}} & \multicolumn{1}{c}{\shortstack{Energy MAE \\ \relax [meV/atom]}} & \multicolumn{1}{c}{\shortstack{Force MAE \\ \relax [eV/\mbox{\normalfont\AA}]}}  \\ \midrule
		$2$ & $87,000$  & $2.54$ & $32.5$ & $0.213$ \\
		$4$ & $235,000$  & $1.92$ & $23.9$ & $0.167$ \\
		$8$ & $529,000$  & $1.65$ & $21.4$ & $0.143$ \\
		$16$ & $1,120,000$  & $\mathbf{1.62}$ & $\mathbf{19.3}$ & $\mathbf{0.142}$ \\ \bottomrule
	\end{tabular}
\end{table}

For further evaluation, we also trained our model for datasets of
locally stable atomic configurations (QM9 dataset and Materials
Project molecule dataset). In addition, we evaluated the applicability
of previous works using our dataset. The results are shown in Appendix
\ref{sec:app:comparison}.

\subsection{Effects of the proposed components of the network}

To investigate the effects of the components in our proposed network
architecture, we systematically removed their corresponding functions
and checked each component effect. The results are presented in Table
\ref{tab:component_result}.

First, the network was run without inputting the tensor values (``w/o
tensor'' row in Table \ref{tab:component_result}). To conduct a fair
test, the number of scalar values was increased to maintain the
original number of parameters in the network.  Then, the network was
run without the node convolution gate (``w/o gate'' row in Table
\ref{tab:component_result}). The number of scalar values was again
increased to offset the reduction in the number of parameters.
Finally, the proposed activation function was replaced by the softplus
function (``Softplus'' row in Table \ref{tab:component_result}).  A
four-layer network without the initial 450,000 iterations was used for
comparison.

\begin{table}[th]
	\caption{Comparison between the baseline and the four-layer network with one removed component. }
	\label{tab:component_result}
	\centering
	\begin{tabular}{cccc}
		\toprule
		 & \multicolumn{1}{c}{\shortstack{Test loss\\ function}} & \multicolumn{1}{c}{\shortstack{Energy MAE \\ \relax [meV/atom]}} & \multicolumn{1}{c}{\shortstack{Force MAE \\ \relax [eV/\mbox{\normalfont\AA}]}}  \\ \midrule
		Original four layers & $\mathbf{1.84}$ & $\mathbf{22.6}$ & $\mathbf{0.161}$ \\
		w/o tensor & $2.15$ & $25.5$ & $0.190$ \\
		w/o gate & $1.99$ & $24.5$ & $0.174$ \\
		Softplus & $1.89$ & $24.1$ & $0.165$ \\ \bottomrule
	\end{tabular}
\end{table}

The largest decrease in accuracy is seen in the case without a tensor
value. The second largest decrease is in the case where the node
convolution gate was not inserted.  Interestingly, the proposed
activation function outperformed the softplus function.

\section{Materials Applications}
\label{sec:applications}
\subsection{Overview}
The universal NNIP should be applicable to arbitrary 3D atomic
configurations with any bond types, crystal/molecular structures, and
element type (up to Ar in this dataset). We have tested various
systems including molecular systems, inorganic crystal structures,
water, and aqueous solutions.

In this section, we used the four-layer version of the neural network
in consideration of the calculation cost of MD simulations. This is
like the embedded-atom potential with embedding applied four times,
and with tensors and vectors propagating inside as well. The same
parameter set is used throughout this section.

\subsection{Intramolecular structure}
We tested the reproducibility of the structures of small C-H
molecules. The bond lengths and bond angles of typical small
hydrocarbon molecules were compared, and the results are listed in
Table \ref{tab:structure}.

Our model can reproduce both the bond lengths and angles with good
accuracy. In particular, a variety of C-C bonding ($sp$, $sp2$, and
$sp3$) is well reproduced. It is noted that ethene forms a planar
structure and that ethane forms a staggered conformation. This
indicates that our model captures the dihedral angle (4-node)
interactions by passing vector and tensor information through the C-C
bond. In addition, we confirmed that benzene forms a planar structure
while cyclohexene forms a chair-type structure, which is a typical
difference in bonding nature between aromaticity and a single bond.

\subsection{Bulk properties of metal and semiconductor}
Metals have delocalized dielectric response, while materials with
bandgap can have exponentially localized response
\cite{QianLQWCYHY08}. Table \ref{tab:structure} shows TeaNet
predictions of Na, Al, and Si. Several crystal structure polymorphs of
the same element were evaluated.

\begin{table}[th]
	\centering
	\caption{Top: Structural accuracy on small hydrocarbon molecules. Bottom: calculated lattice constants and cohesive energies of different phases of Na, Al, and Si. The cohesive energies corresponding to the most stable structure are shown in bold.}
	\begin{tabular}{ccccccc}
		\toprule
		 & \multicolumn{2}{c}{C-C length [\mbox{\normalfont\AA}]} & \multicolumn{2}{c}{C-H length [\mbox{\normalfont\AA}]} & \multicolumn{2}{c}{H-C-C angle [degree]} \\ 
		 & \multicolumn{1}{c}{DFT} & \multicolumn{1}{c}{TeaNet} & \multicolumn{1}{c}{DFT} & \multicolumn{1}{c}{TeaNet} & \multicolumn{1}{c}{DFT} & \multicolumn{1}{c}{TeaNet}  \\ \midrule
		Acetylene (C${}_2$H${}_2$) & $1.21$ & $1.21$ & $1.07$ & $1.06$ & $180^\circ$ & $180^\circ$ \\
		Ethene (C${}_2$H${}_4$) & $1.33$ & $1.34$ & $1.09$ & $1.09$ & $122^\circ$ & $121^\circ$ \\
		Ethane (C${}_2$H${}_6$) & $1.53$ & $1.53$ & $1.10$ & $1.10$ & $112^\circ$ & $112^\circ$ \\
		Benzene (C${}_6$H${}_6$) & $1.40$ & $1.40$ & $1.09$ & $1.09$ & $120^\circ$ & $120^\circ$ \\
		Cyclohexene (C${}_6$H${}_{12}$) & $1.53$ & $1.55$ & $1.10$ & $1.10$ & $110^\circ$ & $110^\circ$ \\ \bottomrule\\
	\end{tabular}
	\bigskip
    \begin{tabular}{llcccc}
		\toprule
		\multicolumn{2}{c}{} & \multicolumn{2}{c}{Lattice constant [\mbox{\normalfont\AA}]} & \multicolumn{2}{c}{Cohesive energy [eV/atom]} \\ 
		\multicolumn{2}{c}{} & \multicolumn{1}{c}{DFT} & \multicolumn{1}{c}{TeaNet} & \multicolumn{1}{c}{DFT} & \multicolumn{1}{c}{TeaNet}  \\ \midrule
		Na & FCC & $5.30$ & $5.39$ & $\mathbf{1.10}$ & $\mathbf{1.16}$ \\
		& BCC & $4.22$ & $4.30$ & $1.09$ & $1.15$ \\
		& Diamond & $7.62$ & $7.29$ & $0.76$ & $0.77$ \\
		Al & FCC & $4.05$ & $4.11$ & $\mathbf{3.42}$ & $\mathbf{3.43}$ \\
		& BCC & $3.23$ & $3.26$ & $3.27$ & $3.38$ \\
		& Diamond & $6.05$ & $6.30$ & $2.79$ & $2.75$ \\
		Si & FCC & $3.91$ & $4.26$ & $3.97$ & $4.43$ \\
		& BCC & $3.17$ & $3.37$ & $3.93$ & $4.40$ \\
		& Diamond & $5.47$ & $5.47$ & $\mathbf{4.64}$ & $\mathbf{4.76}$ \\ \bottomrule
	\end{tabular}
	\label{tab:structure}
\end{table}

\subsection{Amorphous silicon dioxide}
\begin{figure}[th]
    \centering
    \includegraphics[width=0.9\linewidth,trim=00 00 00 00]{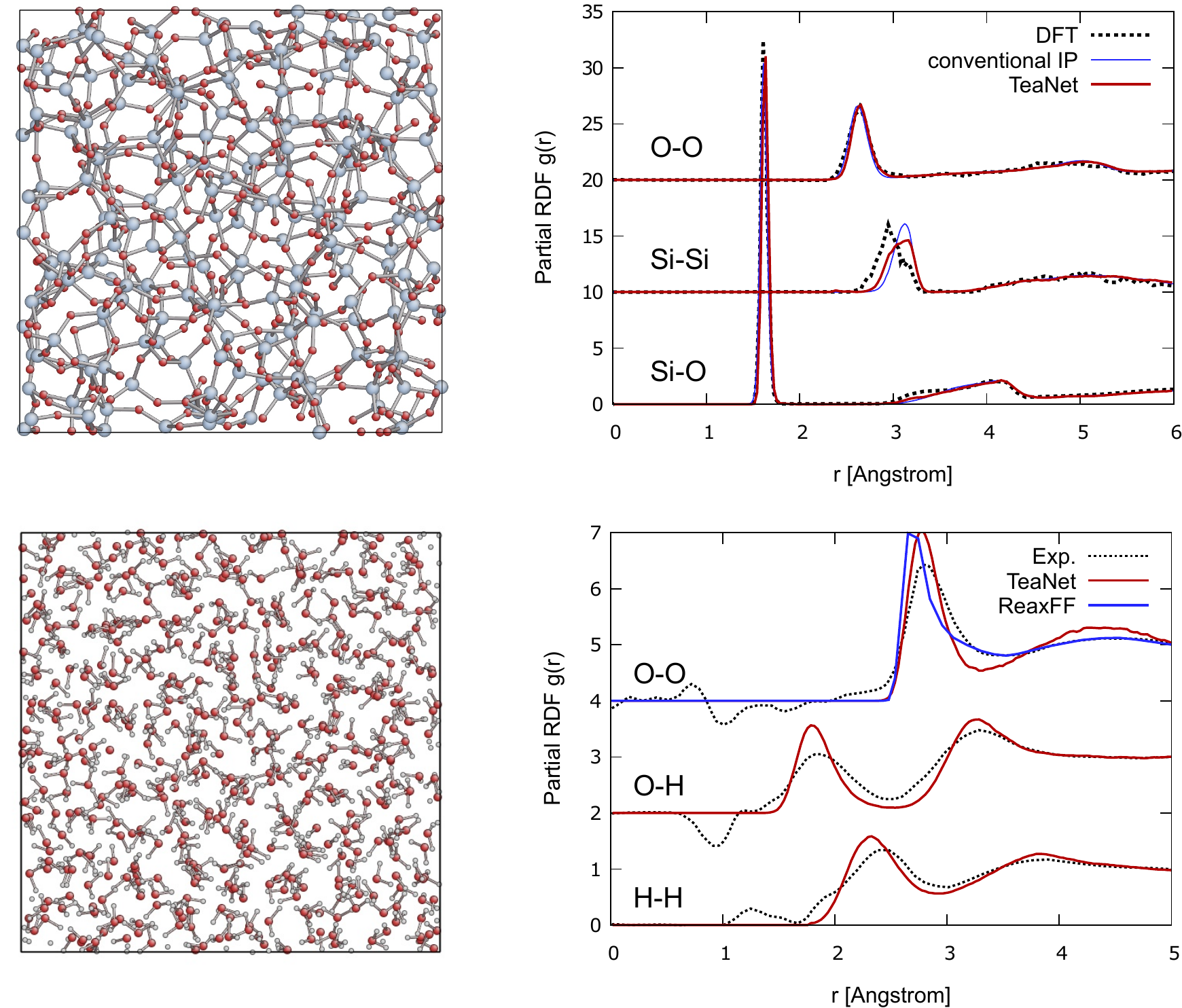}
    \caption{Top left: obtained SiO${}_2$ amorphous structure. Top right: comparison of partial radial distribution function of amorphous SiO${}_2$ with DFT\cite{sarnthein1995model} and conventional IP\cite{doi:10.1063/1.4965863}. Bottom left: snapshot of water. Bottom right: partial radial distribution function of water at 300 K. The experimental data is derived from the merged X-ray and neutron scattering data \cite{soper2013radial}. It is noted that the intramolecular bonds of H${}_2$O (within 1.20 \mbox{\normalfont\AA} for O--H and 1.77 \mbox{\normalfont\AA} for H--H) are not shown. \added{For O--O, ReaxFF potential result \cite{6684cdb1e9924c7aa7c342f6ad206be4} is also shown.}}
	\label{fig:sio2_h2o}
\end{figure}

Since SiO${}_2$ amorphous structure has various bond angles and various coordination numbers, it is treated as a benchmark of the IPs \cite{munetoh2007interatomic, doi:10.1063/1.4965863}. Amorphous SiO${}_2$ configuration including 648 atoms is obtained by a melt-quench process. The obtained structure and the partial radial distribution functions are  shown in Fig. \ref{fig:sio2_h2o}. The result is in good agreement with those of previous studies \cite{sarnthein1995model, doi:10.1063/1.4965863}. Detailed comparison for silica polymorphs \added{with the other IPs \cite{munetoh2007interatomic, 359b1de5761149c79cf6c91071f48994}} are shown in Appendix \ref{sec:app:silica_crystal}.

\subsection{Properties of water}
Water is ubiquitous in chemistry and biochemistry. Atomistic
simulation of polar and protic solvent is, therefore, essential for
chemistry, biochemistry and electrochemistry.  First, the ice (ice Ih)
crystal structure was created. The calculated density of ice at 200 K
was 0.93 g/cm${}^3$. Second, liquid properties were investigated. As
an initial structure, an MD cell having 512 H${}_2$O molecules was
prepared. It was melted at 800 K for 1 ps under NVT ensemble and then
annealed at 300 K and 1 bar for 3 ns under NPT ensemble. The density
of liquid water was 1.00 g/cm${}^3$. These values are in good
agreement with the experimental values (0.92 g/cm${}^3$ at 200 K, 1.00
g/cm${}^3$ at 300 K), and we confirmed that the density of water is
higher than that of ice \cite{haynes2014crc}. Figure
\ref{fig:sio2_h2o} shows a snapshot of TeaNet simulation of a system
of water molecules at 300 K, and the partial radial distribution
function \added{(RDF)} of water predicted by our model compared to the
experiment \cite{soper2013radial}. \added{It is noted that there are
  IPs which can reproduce the liquid water and ice structures. For
  example, the calculated density of liquid water and ice using ReaxFF
  potential \cite{6684cdb1e9924c7aa7c342f6ad206be4} are 1.01
  g/cm${}^3$ and 0.96 g/cm${}^3$, respectively. In addition, O--O
  partial RDF of ReaxFF potential is shown in
  Fig. (\ref{fig:sio2_h2o}).}

Another important property of water is its high dielectric
constant. In MD simulation, the dielectric constant $\epsilon$ can be
calculated from the fluctuation of the total dipole moment by
\cite{neumann1983dipole}

\begin{equation}
\epsilon=1+\frac{4\pi}{3Vk_{\rm B}T}\left(\left<M^2\right>-\left<M\right>^2\right),
\end{equation}
where $M$, $V$, $k_{\rm B}$, and $T$ are the dipole moment, volume,
Boltzmann constant, and temperature, respectively. $\left<\right>$
corresponds to the time average operation.  The dipole moment of a
single H${}_2$O molecule is set to 1.8546 Debye in this simulation.
The calculated dielectric constant was around 52 (Experimental value:
78 at 298 K \cite{haynes2014crc}).

In this simulation, the calculation speed was about 0.14 second/step
for 1536 atoms (512 H${}_2$O molecules) using single NVIDIA Titan V
GPU.

\subsection{Ion dissociation and the Grotthuss proton diffusion mechanism}

\begin{figure}[th]
	\centering
	\includegraphics[width=0.9\linewidth,trim=00 00 00 00]{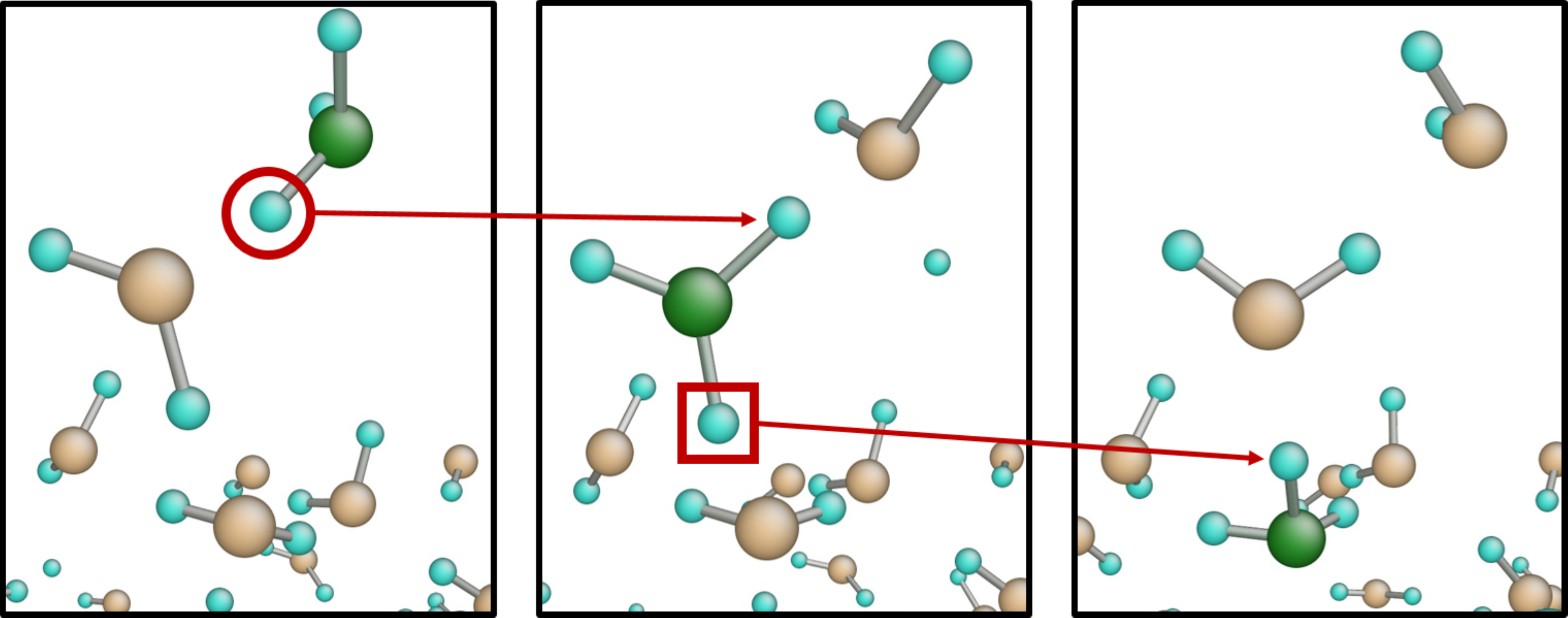}
	\caption{Snapshots of hopping of H between H${}_2$O molecules. H, 2-coordinate O, and 3-coordinate O are shown by blue, yellow, and green spheres, respectively. (Left): In water, H in H${}_2$O and H${}_3$O are oriented to neighboring O atoms. (Left to middle): An H in H${}_3$O hopped to another O. (Middle to right): Another H in the H${}_3$O molecule hopped to the other H${}_2$ O molecule. As a whole, these events were considered as the Grotthuss diffusion of H.}
	\label{fig:grotthuss}
\end{figure}

Next, we investigate ion dissociation, proton transport, and the
Grotthuss mechanism by simulating HCl in H${}_2$O. As a result, the
HCl molecule dissociated and a single Cl atom and H${}_3$O molecule
were created. Here, Cl and H${}_3$O are shown without +/- signs
because the charge deviation effect cannot be extracted
explicitly. After this, occasionally one H atom in the H${}_3$O was
observed to hop to another neighboring O atom, as shown in Figure
\ref{fig:grotthuss}. This proton transfer process, known as the
Grotthuss mechanism, plays an important role in proton diffusion. But
previously there was no bonded IP that can reproduce the Grotthuss
mechanism. \replaced{In TeaNet MD, the calculated effective diffusion
  coefficient of H${}_3$O is $1.5$ \mbox{\normalfont\AA}${}^2$/ps,
  which is in good agreement with the previous DFT study (DFT: $1.3$
  \mbox{\normalfont\AA}${}^2$/ps, experiment: 0.93
  \mbox{\normalfont\AA}${}^2$/ps \cite{boero2005density}). It should
  be noted that ReaxFF potential
  \cite{6684cdb1e9924c7aa7c342f6ad206be4} can reproduce the diffusion
  coefficient (1.0 \mbox{\normalfont\AA}${}^2$/ps).}{In TeaNet MD, the
  calculated effective diffusion coefficient of H${}_3$O is $1.5\times
  10^{-6}$ cm${}^2$/s, which is in good agreement with the previous
  DFT study (DFT: $1.3\times 10^{-6}$ cm${}^2$/s at 300 K
  \cite{boero2005density}).}  The figures focusing on the Cl atom is
shown in Appendix \ref{sec:app:cl_water}.

\section{Conclusion}
In this paper, we provided a unified view of GCN and physics-based
interatomic potentials. Based on the findings, we proposed a new
network model, named the tensor embedded atom network (TeaNet). In
this network, the graph convolution is associated with EAM potential
and the stacked network model is associated with the iterative
electronic total energy relaxation calculation. The Euclidean vectors
and tensor values are incorporated into the model to reproduce the
propagation of orientation-dependent Hamiltonian information. TeaNet
mimics the information flow of nonlinear iterative electronic
relaxations (truncating at 5 iterations at present).  The proposed
model shows great performance for the first 18 elements on the
periodic table (H to Ar) even for highly disordered structures. We
showed that it can reproduce a diverse range of material properties
including C-H molecular structures, metals, amorphous SiO${}_2$,
liquid water and ice.

\subsection*{Data availability}

The raw data required to reproduce these findings are available to download from [Code Ocean (https://codeocean.com/), data available if this manuscript is accepted, or upon request during the review process]. The processed data required to reproduce these findings are available to download from [Code Ocean (https://codeocean.com/), data available if this manuscript is accepted, or upon request during the review process].

\subsection*{Conflict of interest}

Authors declare that there are no conflicts of interest.

\subsection*{Acknowledgments}
JL acknowledges support from the US DOE Office of Nuclear Energy's NEUP Program under Grant No. DE-NE0008751.
ST acknowledges support from a Grant-in-Aid for JSPS Fellows.
We thank Zhe Shi and David Allan Bloore for commenting on the manuscript.

\subsection*{Contributions}
ST performed the research. SI and JL discussed results and
revised the manuscript.


\bibliographystyle{unsrt}
\bibliography{ref.bib}


\clearpage
\appendix
\section*{Appendix}

\subsection{Examples on the weak correlation between bond length and interactions of atoms}
\label{sec:app:length_weak}
\added{With finite radial cutoff distance, using bond length information only sometimes makes it hard to estimate the interactions of atoms. The simple example is ethylene. The rotation of C--C bond is fixed because of pi-bonding. However, with respect to C--C bond rotation, all angles of the chemical bonds which share the same atom do not change. This is interpreted as the dihedral angle interaction which has important role in organic molecules. It is noted that if the cutoff distance is long enough, there can be seen a difference in H--H distance where two hydrogen atoms are connected to the other side of C atoms. However, the change of H--H distance with respect to the rotation of C--C bond is relatively subtle. In addition, in this case, the length-based method should represent the pi-bond interaction as the distance of H--H length, while there is little direct interaction between them.}

\begin{figure} [bhtp]
	\centering
	\includegraphics[width=0.55\linewidth,trim=00 00 00 00]{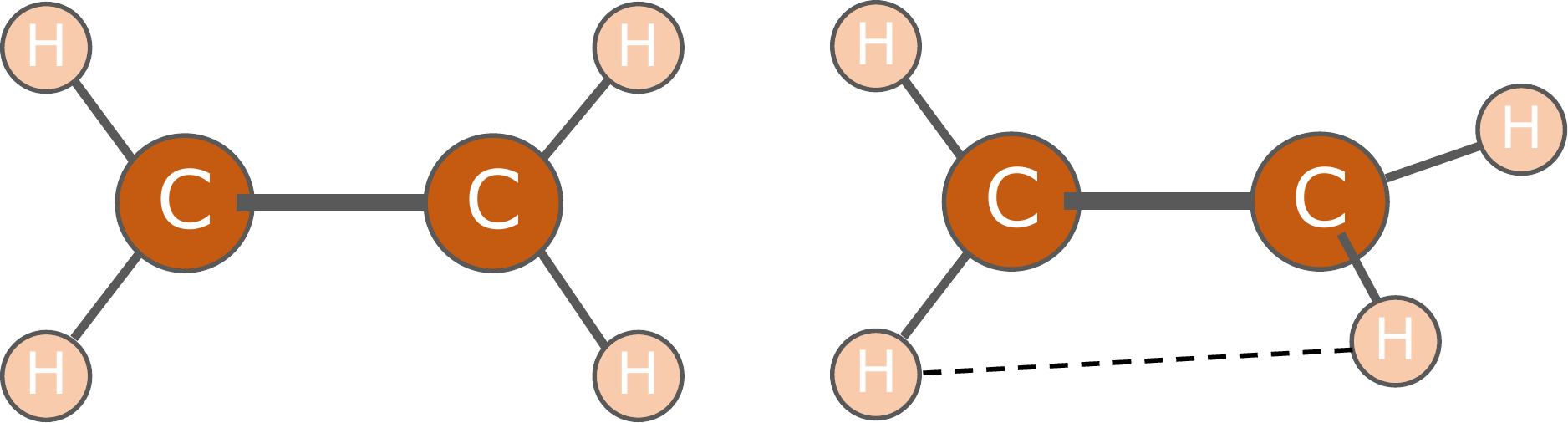}
	\caption{\added{Schematic illustration of the rotation of C--C bond in ethylene. Only H-H distance corresponding opposite site (illustrated by dotted line) is different.}}
	\label{fig:ethylene_rot}
\end{figure}

\added{Another example is small cluster consisting of three atoms arranged in an equilateral triangle. Accounting the nearest neighbor atoms only, the numbers of neighbor atoms are identical to the infinite chain structure. It means that the length-based model with short cutoff distance cannot tell whether the structure is triangle or chain no matter how many the convolution layer is, while their bond angles are quite different. In other words, the length-based model should represent the angle-dependent interaction by the existence of the second nearest neighbor atoms.}

\begin{figure} [bhtp]
	\centering
	\includegraphics[width=0.55\linewidth,trim=00 00 00 00]{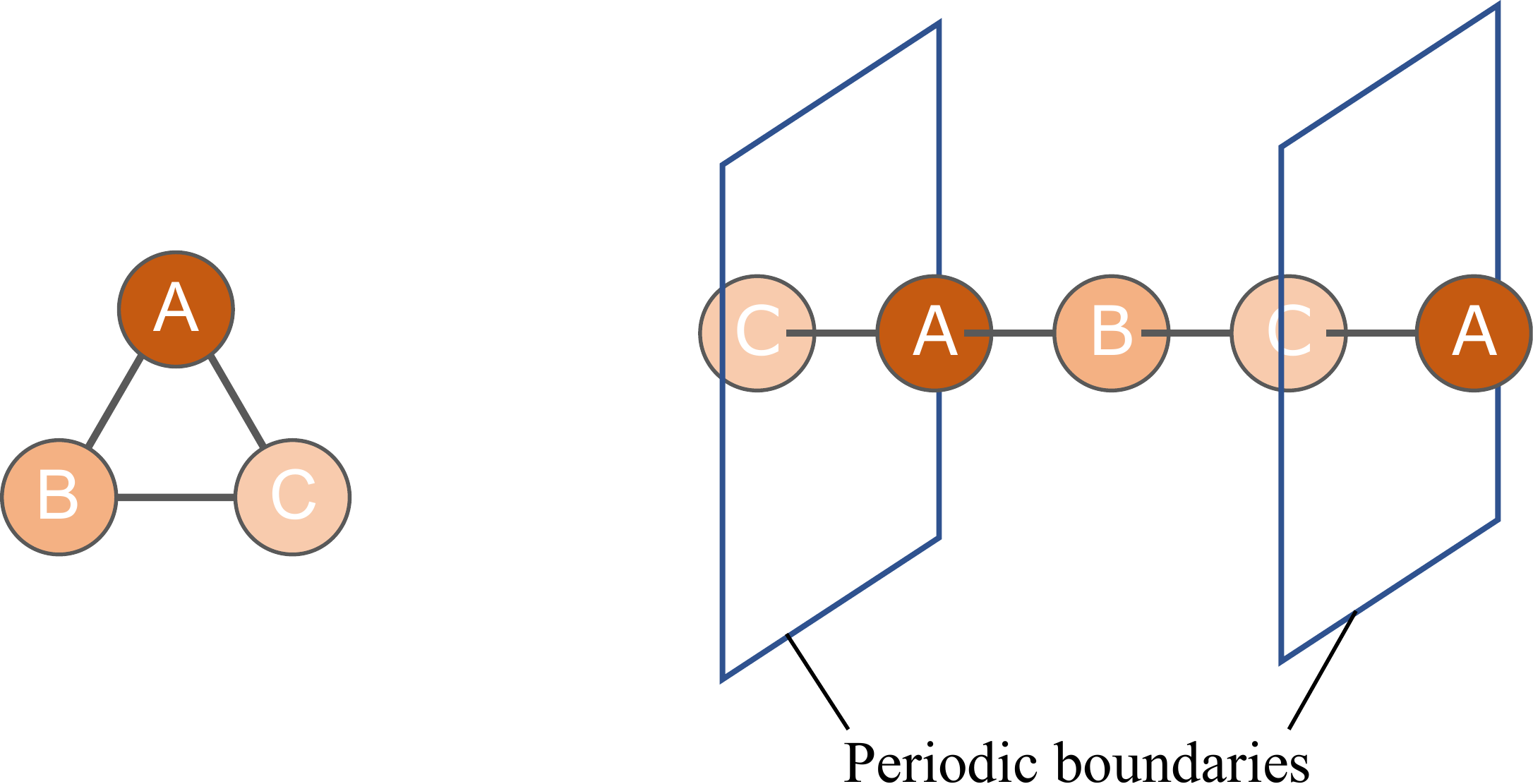}
	\caption{\added{Schematic illustration of triangle cluster and infinite chain. They have the same connectivity.}}
	\label{fig:three_atoms_illustration}
\end{figure}

\clearpage

\subsection{The necessity of Rank-2 tensors and its physical meaning}
\label{sec:app:tensor}

\begin{figure} [bhtp]
	\centering
	\includegraphics[width=0.4\linewidth,trim=00 00 00 00]{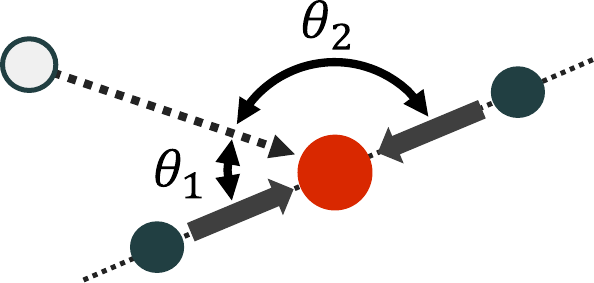}
	\caption{Example of the vanishment of directional information when convoluting with vector values only. If a pair of atoms (shown in dark green circles) having the same properties are located on opposite sides of the center atom (shown in orange circle), any vector values summed at the center atom will vanish. Thus, the angular-dependent interaction between another neighbor atom (shown in white circle) and dark green atoms, corresponding to $\theta_1$ and $\theta_2$, cannot be incorporated in the model.}
	\label{fig:tensor}
\end{figure}

Rank-2 tensors are essential to express the edge-edge interaction through their angle by graph convolution operation. This can be demonstrated in the following example. Let the nodes and edges contain only vector values, and suppose that two edges are connected to a center node, that has point-group symmetry as shown in Fig. \ref{fig:tensor}.
After the convolution, the summed vector values at the node are always $\bf{0}$, and the node loses its directional information. If the third edge is connected to the node, no angle dependence is represented. However, if the second-order tensor values are introduced, the point-group symmetric edge pairs have identical (no sign reversal) tensor values; therefore, the directional information can be accumulated on the node. It should be noted that the vector and tensor values are not merely mathematical tricks but express various physical quantities related to the electronic structure. For example, the local charge deviation is expressed by the electric dipole moment. Since the electron orbit of a $\pi$ bond extends perpendicularly to the bond direction, the dihedral bending is prevented. Polarizability can be expressed by tensor as well. These properties can be naturally expressed using the vector and tensor variables. Higher-order tensor values can also be introduced in the same manner.

\added{The spherical harmonic-based methods \cite{kondor2018clebsch, anderson2019cormorant} also have the ability to represent the higher-order geometric information while holding the rotation and translation invariances from the perspective of SO$(3)$ group.
It should be noted that the tensor product-based representation and the spherical harmonics-based representation are different. The significant difference appears in mirror transformation. Since SO$(3)$ do not hold mirror transformation invariance, the corresponding calculations behave different under mirror-symmetry structure. On the other hand, our method outputs the same value in any O$(n)$ transformations.}

\added{One of this example is outer product of two vectors. Since outer product operation introduces pseudo-vector and breaks mirror symmetry, the tensor product-based method cannot reproduce outer product operation. Instead, Rank-2 rotation tensor which holds O$(n)$ symmetry can be created. This characteristic introduces the desired inductive bias in the field of physics.}

\clearpage

\subsection{Numerical experiments using existing dataset and applicability of other models for our dataset}
\label{sec:app:comparison}
\added{
Although our aim is to reproduce the potential energy of highly disordered atomic configurations, we also evaluated our model for datasets of locally stable atomic configurations.
First, the QM9 dataset was used. Since QM9 contains only stable structures, it is possible to increase accuracy by retraining.
We retrained the four-layer version of the network with the stochastic gradient descent (SGD) optimizer while gradually decreasing the learning rate. The mean squared error of the energy was used as the loss. In this case, we use the original QM9 validation dataset as the test dataset.
The MAE of the energy was 13 meV per molecule (1.2 meV/atom) among the QM9 validation dataset. This is similar to the current top scores (14 meV \cite{schutt2017schnet}, 8 meV \cite{doi:10.1021/acs.jctc.9b00181}), and the other methods (19-130 meV) \cite{gilmer2017neural}.
It is noted that the error of the dataset with locally stable structures is one magnitude smaller than that of highly disordered structures shown in Table \ref{tab:layer_result}.
}

\added{
Second, the Materials Project molecule dataset, which consists of elements in the first three rows of the periodic table, was used. We recalculated the energy of the dataset by DFT to adjust the difference in the method of DFT.
We trained the network in the same way as with QM9. The resulting MAE of the energy was 3.1 meV/atom. Our model well succeeds in estimating the energy of locally stable atomic configurations. It is noted that our model does not require the bond types as the input and that we use a relatively short cutoff distances (6 \mbox{\normalfont\AA}).
}

\added{
For further comparison, we discuss the applicability of the other models for our highly disordered dataset.
First, we would like to note that symmetry function-based methods (BPNN \cite{PhysRevLett.98.146401} and its derivations) is not suitable for this task. Since symmetry function explicitly treats the three-body term of each element, the number of parameters increases dramatically by increasing the number of elements in the dataset. This behavior makes it hard to train the model.}

\added{
On the other hand, GNN-based model can be applied to this dataset. We use SchNet \cite{schutt2017schnet} as the current length-based milestone method. To focus on the reproducibility of dynamics properties, we mainly focus on the atomic forces. We use SchNetPack for the evaluation. The specified parameters are below. To align the conditions, we set the number of layer to 4 (original model: 3 and 6) and the cutoff distance to 6.0 \mbox{\normalfont\AA}. We set the parameter $\rho$ (weight of losses) for the energy and the force to 0.001 and 0.999, respectively. The result is shown in Table \ref{tab:comparison_schnet}.
}

\begin{table} [bhtp]
    \centering
	\caption{Model comparison using proposed highly disordered dataset.}
    \label{tab:comparison_schnet}
	\begin{tabular}{lccc}
		\toprule
		 & \multicolumn{1}{c}{\# params} &  \multicolumn{1}{c}{\shortstack{Energy MAE \\ \relax [meV/atom]}} & \multicolumn{1}{c}{\shortstack{Force MAE \\ \relax [eV/\mbox{\normalfont\AA}]}}  \\ \midrule
		SchNet \cite{schutt2017schnet} & $310,000$ & $29.7$ & $0.638$ \tabularnewline
		ours & $235,000$ & $23.9$ & $0.167$ \tabularnewline \bottomrule
	\end{tabular}
\end{table}

\added{
There is a large difference on estimating atomic forces between SchNet and ours. It should be noted that the structures of this dataset is not limited to specific molecular systems. It means the model is requested to reproduce the properties of structures which is not supposed to exist in the training dataset. This task will be further difficult as compared to the existing dynamics benchmarks using specific molecular systems.
}

\clearpage

\subsection{Silica polymorphs reproducibility}
\label{sec:app:silica_crystal}
\begin{figure} [bhtp]
    \centering
    \includegraphics[width=0.7\linewidth,trim=00 00 00 00]{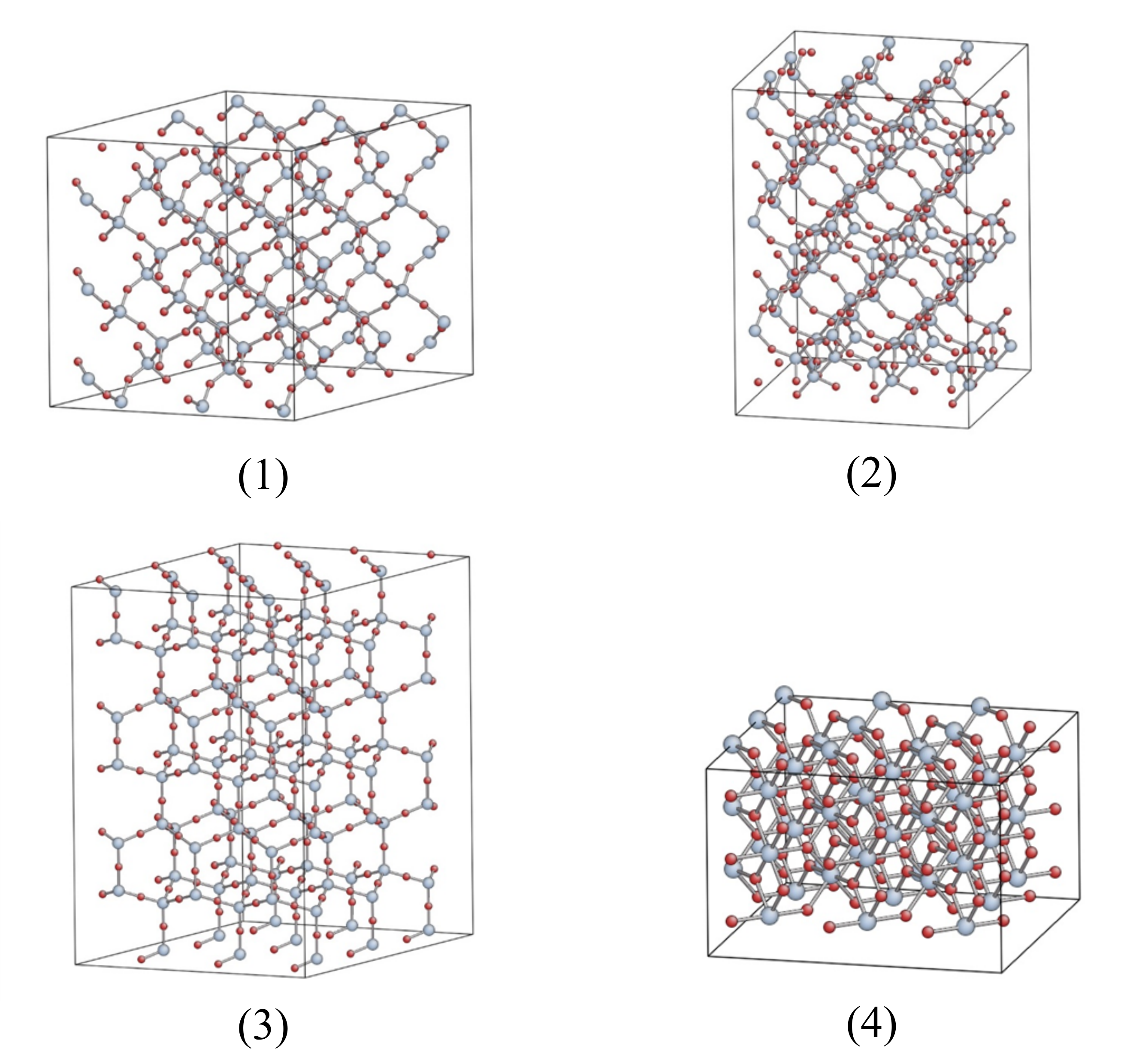}
	\caption{Illustration of SiO2 crystals (1) $\alpha$-quartz. (2) $\alpha$-cristobalite. (3) $\beta$-tridymite. (4) stishovite. The optimized structure using TeaNet are shown. It is noted that Si and O atoms in stishovite have 6 neighboring atoms and 3 neighboring atoms, respectively.}
	\label{fig:sio2_polymorphs}
\end{figure}

\added{Additional experiments for the reproducibility of silica polymorphs ($\alpha$-quartz, $\alpha$-cristobalite, $\beta$-tridymite, stishovite) are carried out. The result is shown in table \ref{tab:silica_crystals}. The snapshots are shown in Fig. \ref{fig:sio2_polymorphs}. Overall, our model well reproduces the silica polymorphs including the difference of the energies and the densities. In addition, our model well estimates the difference of the energy of stishovite, which Tersoff-type potential estimates two times larger. In stishovite crystal structure, one Si atom is connected to 6 O atoms and one O atom is connected to 3 Si atoms. It means that the local environment of each atom is far from the usual tetrahedral silica crystal structures. It indicates that our model is robust for the change of local environments of atoms.}

\begin{table} [bhtp]
    \centering
	\caption{Calculated cohesive energy, relative energy to $\alpha$-quartz, and density of silica polymorphs. a-Q, a-C, and b-T correspond to $\alpha$-quartz, $\alpha$-cristobalite, $\beta$-tridymite, respectively.}
    \label{tab:silica_crystals}
    \begin{tabular}{llccc}
		\toprule
		 & \shortstack{Crystal \\ \relax} & \shortstack{Cohesive energy \\ \relax [eV/atom]} & \shortstack{Relative energy \\ \relax [eV/atom]} & \shortstack{Density \\ \relax [g/cm${}^3$]} \tabularnewline  \midrule
        DFT \cite{demuth1999polymorphism} & a-Q & $7.942$ &  & $2.48$ \tabularnewline
                   & a-C & $7.953$ & $-0.011$ & $2.13$ \tabularnewline
                   & b-T & $7.950$ & $-0.008$ & $2.06$ \tabularnewline
                   & stishovite & $7.730$ & $0.212$ & $4.11$ \tabularnewline
        Tersoff \cite{munetoh2007interatomic} & a-Q & $6.698$ & & $2.42$ \tabularnewline
                       & a-C & $6.697$ & $0.001$ & $2.16$ \tabularnewline
                       & b-T & $6.696$ & $0.002$ & $2.08$ \tabularnewline
                       & stishovite & $6.196$ & $0.502$ & $3.89$ \tabularnewline
        ReaxFF \cite{359b1de5761149c79cf6c91071f48994} & a-Q & - & & $2.55$ \tabularnewline
                       & a-C & - & $0.001$ & $2.22$ \tabularnewline
                       & b-T & - & $-0.006$ & $2.09$ \tabularnewline
                       & stishovite & - & $0.279$ & $4.29$ \tabularnewline
        Ours & a-Q & $6.720$ & & $2.44$ \tabularnewline
             & a-C & $6.717$ & $0.003$ & $2.19$ \tabularnewline
             & b-T & $6.711$ & $0.009$ & $1.92$ \tabularnewline
             & stishovite & $6.466$ & $0.254$ & $4.12$ \tabularnewline \bottomrule
	\end{tabular}
\end{table}

\clearpage

\subsection{Cl atom observation in water}
\label{sec:app:cl_water}
\begin{figure} [bhtp]
    \centering
    \includegraphics[width=0.75\linewidth,trim=00 00 00 00]{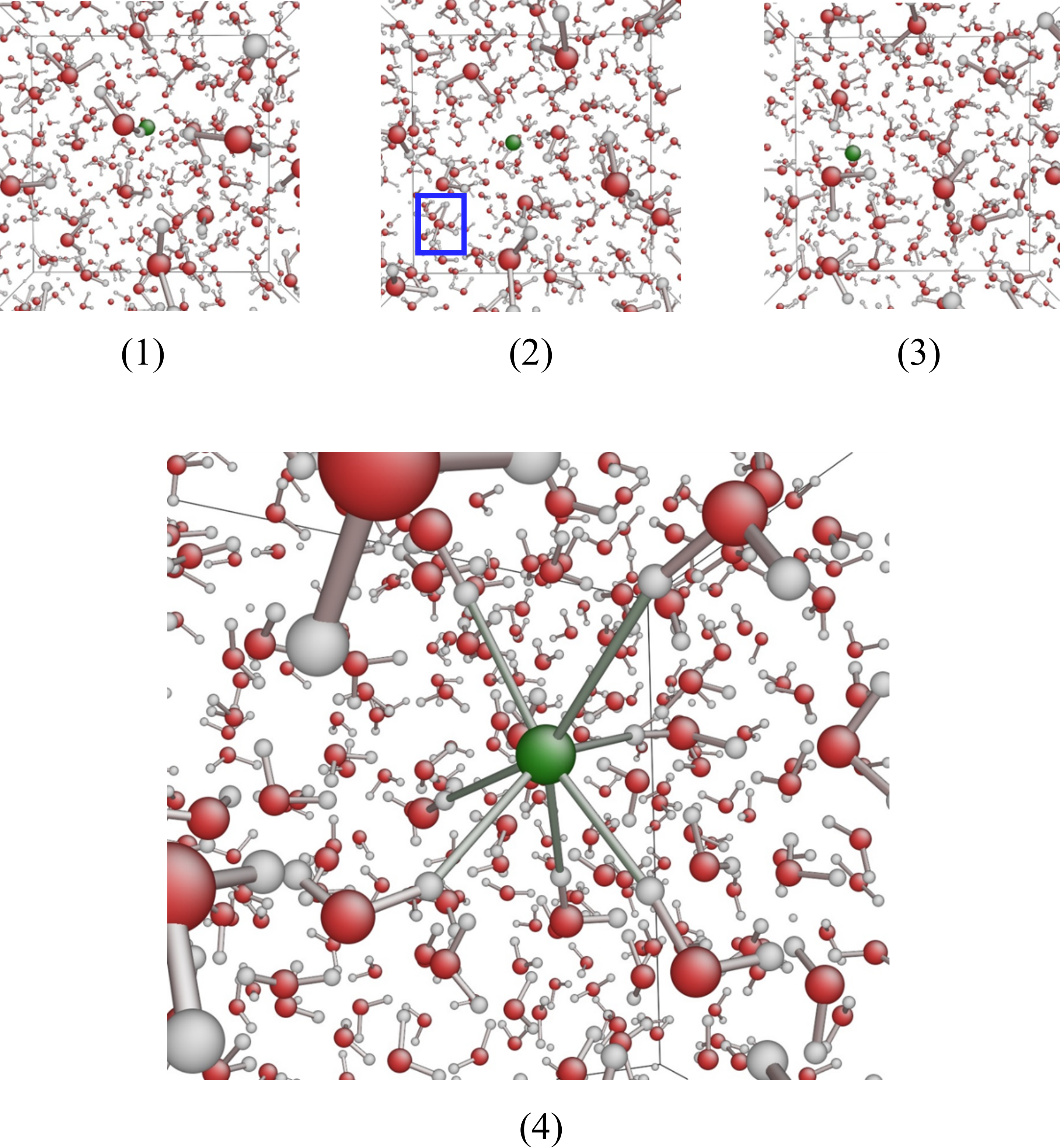}
	\caption{(1), (2), and (3): Snapshots of water in which HCl molecule was inserted. The green sphere corresponds to Cl atom. H3O molecule can be found in the blue box shown in (2). (4): Close snapshot of (3). H-Cl long-range bonds are also shown (visual cutoff distance for H-Cl was set to 3 \mbox{\normalfont\AA}).}
	\label{fig:cl_trajectory}
\end{figure}

In the simulation of ion dissociation and proton diffusion of water, one HCl molecule was added into H${}_2$O. In this section, the behavior of Cl atom was observed. The snapshots are shown in Fig. \ref{fig:cl_trajectory}. As the HCl molecule dissociated in the water, the individual Cl atom was observed during the MD simulation. The interaction of Cl atom and surrounding water molecules was also shown. Although they are not bonded strictly, H atoms in the surrounding water molecule tend to get closer to the Cl atom. This is in good agreement with the picture of anions in water. It should be noted that those effects were reproduced without preparing any explicit water-Cl DFT simulations in advance.

\clearpage

\subsection{Details of dataset}
\label{sec:app:dataset_detail}
The details of the test dataset is shown. Since it was made by randomly selecting from the entire dataset, it can be considered to reflect the trend of the entire dataset.

\begin{figure}
	\centering
	\includegraphics[width=0.93\linewidth,trim=00 00 00 00]{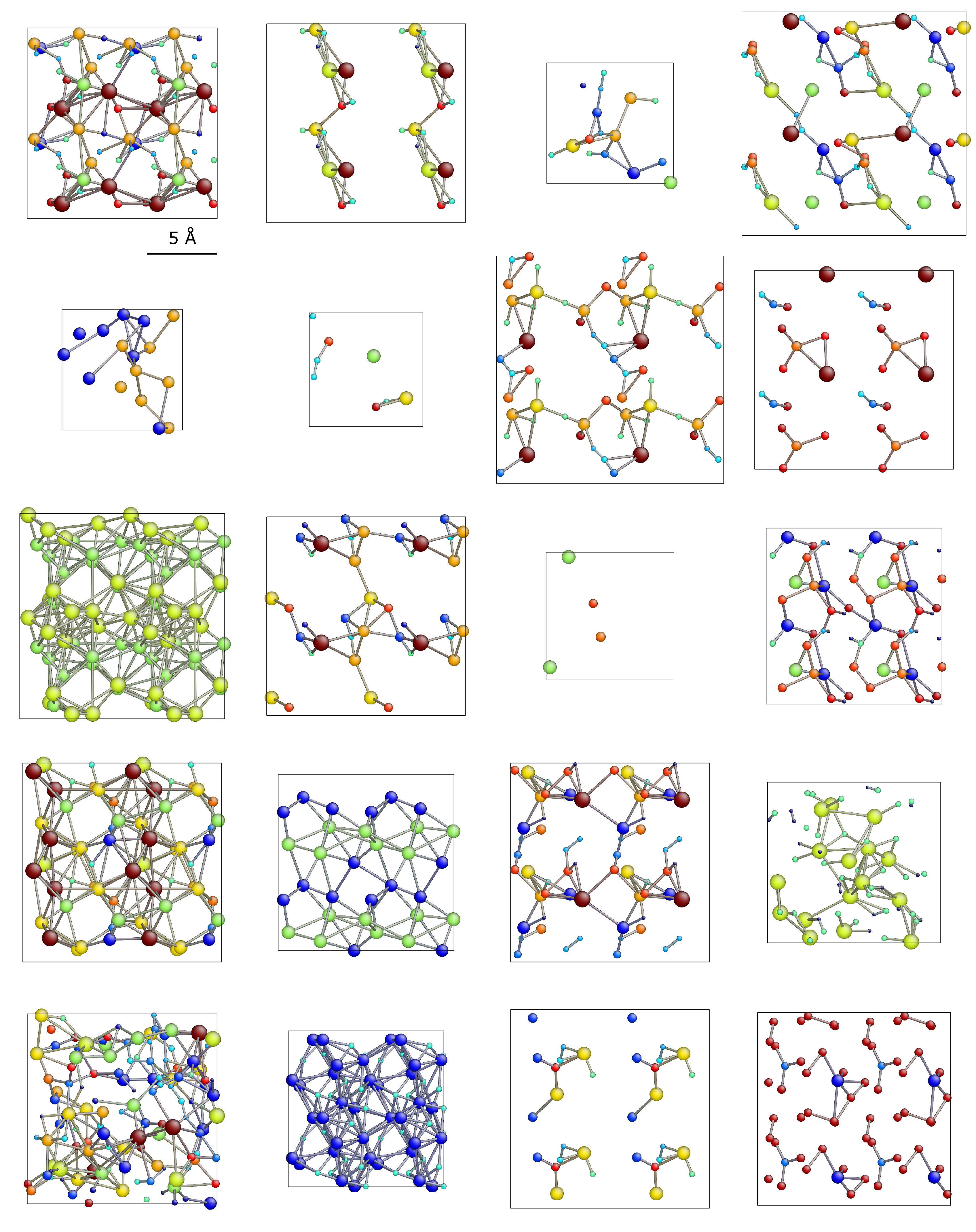}
	\caption{The structures of the first 20 samples in table \ref{tab:app1} are shown (order: top left to right). The colors of the atoms correspond to the element number (H: blue, Ar: red). It is noted that the structures with small box are drawn at a size of $2\times 2$. See table \ref{tab:app1} for the details of component.}
	\label{fig:app_atoms}
\end{figure}

Table \ref{tab:app1} shows the amount of each element in the test dataset. The calculated energy is also shown. The structures of the first 20 samples in table \ref{tab:app1} are shown in Fig. \ref{fig:app_atoms}. Table \ref{tab:app2} shows the number of pairs in the test dataset. As described in the main text, the dataset consists of highly disordered structures.

\begin{center}
\begin{longtable}{rrrrrrrrrrrrrrrrrrD{.}{.}{2}D{.}{.}{2}}
	\caption{Content of the test dataset. The number of each element in the test dataset is shown. $E$ corresponds to the total energy of the system calculated by DFT. The zero point of the energy is defined as the sum of the energies of atoms separated in a vacuum. The unit of energy is eV.}
	\label{tab:app1}
	\\
		\toprule
		\multicolumn{1}{c}{H} & \multicolumn{1}{c}{He} & \multicolumn{1}{c}{Li} & \multicolumn{1}{c}{Be} & \multicolumn{1}{c}{B} & \multicolumn{1}{c}{C} & \multicolumn{1}{c}{N} & \multicolumn{1}{c}{O} & \multicolumn{1}{c}{F} & \multicolumn{1}{c}{Ne} & \multicolumn{1}{c}{Na} & \multicolumn{1}{c}{Mg} & \multicolumn{1}{c}{Al} & \multicolumn{1}{c}{Si} & \multicolumn{1}{c}{P} & \multicolumn{1}{c}{S} & \multicolumn{1}{c}{Cl} & \multicolumn{1}{c}{Ar} & \multicolumn{1}{c}{$E$} & \multicolumn{1}{c}{ $E$ /atom} \\
		\midrule
		\endfirsthead
		\toprule
		\multicolumn{1}{c}{H} & \multicolumn{1}{c}{He} & \multicolumn{1}{c}{Li} & \multicolumn{1}{c}{Be} & \multicolumn{1}{c}{B} & \multicolumn{1}{c}{C} & \multicolumn{1}{c}{N} & \multicolumn{1}{c}{O} & \multicolumn{1}{c}{F} & \multicolumn{1}{c}{Ne} & \multicolumn{1}{c}{Na} & \multicolumn{1}{c}{Mg} & \multicolumn{1}{c}{Al} & \multicolumn{1}{c}{Si} & \multicolumn{1}{c}{P} & \multicolumn{1}{c}{S} & \multicolumn{1}{c}{Cl} & \multicolumn{1}{c}{Ar} & \multicolumn{1}{c}{$E$} & \multicolumn{1}{c}{ $E$ /atom} \\ 
		\midrule
		\endhead
0&1&1&0&0&3&0&1&3&1&0&0&3&0&0&1&2&2&-32.57 & -1.81 \\
1&0&0&0&0&0&0&2&1&0&1&1&0&0&0&1&0&1&-22.14 & -2.77 \\
0&1&1&1&2&2&0&2&2&1&0&1&2&0&1&0&0&0&-58.83 & -3.68 \\
0&1&1&1&0&1&2&1&1&1&1&1&0&1&1&1&1&1&-53.44 & -3.34 \\
0&0&8&0&0&0&0&0&0&0&0&0&8&0&0&0&0&0&-30.25 & -1.89 \\
0&0&0&0&0&0&3&1&0&1&0&1&0&0&1&0&1&0&-18.42 & -2.30 \\
0&0&0&0&1&1&2&0&4&0&0&1&2&1&2&0&1&1&-64.72 & -4.05 \\
0&0&0&0&1&0&1&0&0&0&0&0&0&1&0&2&2&1&-22.55 & -2.82 \\
0&0&0&0&0&0&0&0&0&9&9&0&0&0&0&0&0&0&0.97 & 0.05 \\
0&1&0&2&0&0&1&0&1&0&0&1&2&0&1&0&0&1&-25.22 & -2.52 \\
0&0&0&0&0&0&0&0&0&2&0&0&0&1&1&0&0&0&-2.54 & -0.63 \\
3&0&2&0&0&1&0&0&2&1&0&0&0&1&3&1&2&0&-47.47 & -2.97 \\
0&0&1&0&1&0&0&1&1&2&1&4&1&1&0&0&0&3&-26.56 & -1.66 \\
0&0&4&0&0&0&0&0&0&4&0&0&0&0&0&0&0&0&-2.99 & -0.37 \\
2&0&2&0&1&3&0&1&0&0&0&1&1&1&3&0&0&1&-55.86 & -3.49 \\
16&0&0&0&0&0&0&0&32&0&16&0&0&0&0&0&0&0&-197.82 & -3.09 \\
9&6&10&12&7&12&9&3&7&9&6&10&5&3&3&7&6&4&-394.97 & -3.09 \\
0&0&9&0&0&0&0&9&0&0&0&0&0&0&0&0&0&0&-62.79 & -3.49 \\
0&0&0&2&0&1&1&0&1&0&0&2&0&0&0&1&0&0&-23.62 & -2.95 \\
0&0&1&0&1&0&0&0&0&0&0&0&0&0&0&0&14&0&-28.70 & -1.79 \\
0&0&8&0&0&0&0&0&8&0&0&0&0&0&0&0&0&0&-58.89 & -3.68 \\
2&0&3&1&1&1&1&0&2&0&0&0&3&0&1&1&0&0&-48.22 & -3.01 \\
1&2&2&0&0&1&1&1&0&0&0&3&1&0&0&0&2&2&-27.45 & -1.72 \\
2&1&2&1&0&0&0&0&3&3&0&0&2&0&1&1&0&0&-41.19 & -2.57 \\
0&0&0&0&0&1&0&0&0&0&0&0&0&0&0&0&0&7&1.25 & 0.16 \\
0&0&0&0&0&0&0&0&0&1&0&15&0&0&0&0&0&0&-0.22 & -0.01 \\
0&0&16&0&0&0&0&0&0&0&0&0&0&0&0&0&0&0&-28.89 & -1.81 \\
10&0&0&0&0&12&2&0&0&0&0&0&0&0&0&0&0&0&-117.52 & -4.90 \\
1&1&0&2&1&0&1&2&1&0&2&1&1&1&1&0&1&0&-56.39 & -3.52 \\
1&1&0&0&1&0&1&1&1&0&0&0&0&1&2&0&1&0&-36.41 & -3.64 \\
1&1&1&0&2&1&0&0&2&1&2&1&0&2&1&1&1&1&-45.35 & -2.52 \\
0&3&1&0&0&0&1&3&0&1&0&2&0&0&1&2&2&0&-36.98 & -2.31 \\
2&0&2&2&0&2&0&0&1&2&2&0&2&1&2&0&0&2&-34.90 & -1.74 \\
0&0&0&0&0&0&0&0&0&9&0&0&0&0&9&0&0&0&-20.54 & -1.14 \\
0&0&0&0&1&0&1&1&0&1&0&2&0&1&0&0&1&0&-22.94 & -2.87 \\
4&1&2&2&0&0&1&1&1&0&1&1&1&0&1&0&0&0&-35.49 & -2.22 \\
1&1&0&1&2&1&1&0&0&1&2&0&1&1&1&2&0&3&-32.36 & -1.80 \\
1&0&1&0&0&2&1&0&0&0&1&0&0&0&0&2&0&0&-25.29 & -3.16 \\
4&5&2&2&4&5&3&5&1&6&2&4&1&2&2&9&5&2&-220.63 & -3.45 \\
0&2&1&1&1&1&0&0&1&1&0&1&0&1&0&0&0&0&-23.16 & -2.32 \\
2&0&0&0&0&0&0&1&0&2&0&0&0&0&0&0&2&1&-10.70 & -1.34 \\
0&0&0&0&2&0&0&0&0&0&0&0&0&0&0&0&0&0&-3.60 & -1.80 \\
1&2&0&0&0&2&2&2&1&1&0&1&1&0&1&0&2&0&-60.44 & -3.78 \\
1&2&0&0&0&1&0&0&0&1&0&0&1&0&1&1&0&0&-11.70 & -1.46 \\
0&0&0&0&0&0&0&0&0&0&0&0&0&0&0&10&10&0&-41.19 & -2.06 \\
0&0&15&0&0&0&0&0&0&0&1&0&0&0&0&0&0&0&-31.25 & -1.95 \\
0&0&2&2&0&1&1&0&0&0&0&2&3&0&0&1&3&1&-44.95 & -2.81 \\
0&0&0&0&4&0&0&0&0&0&4&0&0&0&0&0&0&0&-21.54 & -2.69 \\
0&0&0&0&1&0&1&1&0&0&1&0&1&1&0&1&1&0&-26.63 & -3.33 \\
0&0&0&0&0&0&0&0&0&8&0&8&0&0&0&0&0&0&-5.74 & -0.36 \\
1&0&0&0&1&0&1&0&0&1&0&0&2&1&0&0&1&0&-10.77 & -1.35 \\
1&0&0&1&0&2&0&0&1&0&0&0&1&0&1&0&0&1&-25.65 & -3.21 \\
1&1&0&1&0&1&2&1&2&0&0&1&0&1&0&0&0&1&-46.32 & -3.86 \\
5&5&4&4&3&3&3&6&3&4&4&1&5&1&0&3&4&6&-206.37 & -3.22 \\
0&0&0&0&0&16&0&0&0&0&0&0&0&0&0&0&0&0&-22.43 & -1.40 \\
0&0&0&0&0&0&0&0&0&0&4&0&0&4&0&0&0&0&-16.87 & -2.11 \\
6&0&10&0&0&0&0&0&0&0&0&0&0&0&4&0&0&0&-36.95 & -1.85 \\
0&0&0&0&1&0&1&0&0&0&0&0&0&2&0&0&0&0&-13.41 & -3.35 \\
2&0&1&0&1&1&1&0&1&0&1&0&1&1&0&1&1&0&-34.68 & -2.89 \\
0&0&0&0&0&0&0&0&0&0&0&0&9&0&0&0&9&0&-45.46 & -2.53 \\
0&0&0&0&0&0&8&0&0&8&0&0&0&0&0&0&0&0&-41.27 & -2.58 \\
0&2&2&2&1&0&2&1&1&2&0&0&1&1&0&0&1&0&-43.28 & -2.71 \\
0&0&0&0&8&0&0&0&8&0&0&0&0&0&0&0&0&0&-80.56 & -5.03 \\
0&1&2&1&0&0&0&2&0&0&0&0&0&0&0&2&0&0&-24.73 & -3.09 \\
0&0&0&0&0&0&0&0&0&1&0&0&0&0&0&1&1&1&-1.31 & -0.33 \\
0&2&1&0&2&1&0&3&1&1&0&0&2&1&1&1&2&2&-61.66 & -3.08 \\
0&0&0&0&0&0&0&1&0&0&0&0&0&2&0&0&0&1&-10.50 & -2.62 \\
0&0&0&9&0&0&9&0&0&0&0&0&0&0&0&0&0&0&-71.42 & -3.97 \\
2&0&0&0&0&1&1&1&0&0&1&0&1&0&2&2&1&4&-42.65 & -2.67 \\
0&3&1&3&0&0&0&0&1&0&0&2&2&0&2&1&1&0&-39.80 & -2.49 \\
1&0&1&1&0&0&0&1&0&0&2&0&0&1&0&1&0&0&-22.08 & -2.76 \\
5&6&7&10&8&12&14&2&9&10&5&6&3&8&6&9&3&5&-427.13 & -3.34 \\
0&0&0&0&0&0&0&0&1&0&0&6&0&0&0&0&1&0&-15.13 & -1.89 \\
0&0&0&0&16&0&0&0&0&0&0&0&0&0&0&0&0&0&-88.14 & -5.51 \\
2&1&1&1&0&1&1&0&2&1&3&0&0&1&0&1&0&1&-39.07 & -2.44 \\
0&0&0&1&1&0&0&2&2&0&1&1&1&0&3&0&0&0&-42.05 & -3.50 \\
1&1&2&2&3&0&0&0&0&0&1&1&0&0&0&2&2&1&-39.07 & -2.44 \\
0&1&1&1&1&0&0&3&0&1&0&0&1&1&0&0&0&0&-32.15 & -3.22 \\
8&0&0&0&0&8&0&0&0&0&0&0&0&0&0&0&0&0&-71.87 & -4.49 \\
1&0&1&1&0&3&5&0&1&0&1&1&1&1&1&1&1&1&-66.54 & -3.33 \\
1&1&3&1&0&0&1&2&0&1&1&0&0&2&0&1&1&1&-46.73 & -2.92 \\
0&1&0&1&0&2&1&1&0&0&1&1&0&0&0&2&1&1&-38.19 & -3.18 \\
0&2&1&1&1&1&1&0&2&0&0&1&1&1&0&1&0&3&-46.10 & -2.88 \\
0&0&8&0&16&0&8&16&8&8&0&0&16&0&32&8&8&0&-461.06 & -3.60 \\
0&0&0&0&0&0&0&0&0&0&0&0&15&0&0&1&0&0&-44.16 & -2.76 \\
11&8&8&6&8&11&6&3&5&12&9&6&6&6&6&4&6&7&-374.90 & -2.93 \\
0&0&0&0&0&0&0&0&0&0&0&0&16&0&0&0&0&0&-39.70 & -2.48 \\
6&0&0&0&0&8&2&0&0&0&0&0&0&0&0&0&0&0&-75.53 & -4.72 \\
1&0&0&1&0&1&1&0&2&0&0&0&2&3&1&2&0&2&-56.83 & -3.55 \\
1&1&0&0&0&0&1&2&2&1&1&0&0&0&0&2&5&0&-32.44 & -2.03 \\
0&0&1&1&1&0&1&0&0&1&2&0&0&0&0&0&0&1&-7.19 & -0.90 \\
0&0&0&0&0&0&8&0&0&0&0&0&0&0&0&0&0&0&-40.96 & -5.12 \\
0&2&0&1&0&1&1&1&0&1&0&1&2&1&1&0&4&0&-37.51 & -2.34 \\
0&0&0&10&0&10&0&0&0&0&0&0&0&0&0&0&0&0&-93.71 & -4.69 \\
1&2&3&1&0&1&1&0&0&2&1&1&0&2&0&2&2&1&-44.09 & -2.20 \\
4&0&0&0&0&4&2&2&0&0&0&0&0&0&0&0&0&0&-51.94 & -4.33 \\
0&0&0&1&2&0&3&0&1&0&0&1&0&1&2&1&0&0&-53.71 & -4.48 \\
0&0&0&0&0&0&0&0&0&4&0&0&0&4&0&0&0&0&-11.15 & -1.39 \\
0&1&0&0&0&2&1&3&1&2&0&3&0&1&0&0&0&2&-50.47 & -3.15 \\
0&0&8&0&0&0&0&8&0&0&0&0&0&0&0&0&0&0&-57.67 & -3.60 \\
2&2&0&1&2&1&0&0&0&1&1&0&1&2&0&0&1&2&-29.53 & -1.85 \\
14&7&4&0&2&5&13&8&8&7&5&10&5&9&11&7&7&6&-378.64 & -2.96 \\
0&0&0&0&0&9&0&0&0&0&0&0&0&0&0&9&0&0&-74.44 & -4.14 \\
0&1&0&0&0&0&0&0&7&0&0&0&0&0&0&0&0&0&-6.89 & -0.86 \\
0&0&0&0&0&0&9&0&0&0&9&0&0&0&0&0&0&0&-46.17 & -2.56 \\
4&7&7&4&4&4&7&3&0&5&2&3&3&3&2&2&3&1&-204.67 & -3.20 \\
0&0&0&0&0&0&0&1&0&0&0&0&0&0&0&0&1&6&-2.18 & -0.27 \\
0&4&0&0&0&0&0&0&0&0&0&0&0&0&4&4&0&4&-19.82 & -1.24 \\
3&3&1&1&1&2&2&3&3&33&2&3&1&1&2&0&0&3&-92.94 & -1.45 \\
1&3&1&2&0&2&2&1&1&0&1&1&0&1&0&2&0&2&-51.51 & -2.58 \\
0&0&0&1&2&1&2&0&0&0&2&2&2&3&1&0&0&0&-59.71 & -3.73 \\
16&8&16&0&0&0&8&8&0&24&0&0&24&8&0&0&8&8&-266.13 & -2.08 \\
2&2&1&1&0&1&0&1&0&0&0&1&1&0&3&2&0&1&-45.98 & -2.87 \\
0&2&0&0&0&0&0&0&0&0&0&0&0&0&16&0&0&0&-44.08 & -2.45 \\
0&1&0&0&2&0&1&0&1&2&0&2&0&1&2&0&0&0&-24.05 & -2.00 \\
0&0&0&0&0&0&0&0&0&0&0&0&0&0&0&16&0&0&-47.83 & -2.99 \\
8&8&2&8&8&6&10&2&7&8&7&3&11&6&5&9&14&6&-383.19 & -2.99 \\
0&0&0&0&8&0&0&0&0&0&0&0&0&0&0&8&0&0&-76.73 & -4.80 \\
0&0&2&1&3&1&0&2&1&1&0&0&0&1&1&1&0&2&-59.94 & -3.75 \\
0&0&0&0&7&0&1&0&0&0&0&0&0&0&0&0&0&0&-33.89 & -4.24 \\
1&0&3&2&1&1&1&0&1&1&0&0&0&0&0&0&1&0&-29.74 & -2.48 \\
0&1&0&0&0&0&0&0&0&6&0&0&0&0&1&0&0&0&-0.06 & -0.01 \\
7&0&0&0&0&7&0&1&0&0&0&0&0&0&0&1&3&0&-74.36 & -3.91 \\
1&0&1&2&2&1&1&0&1&0&1&1&2&2&1&1&1&2&-54.96 & -2.75 \\
0&8&0&0&0&0&0&0&0&0&0&0&0&0&0&8&0&0&-22.66 & -1.42 \\
0&0&0&0&2&1&1&1&0&1&0&2&1&1&0&0&1&1&-31.12 & -2.59 \\
0&0&1&0&0&0&0&1&0&0&1&0&0&0&0&0&0&0&-5.95 & -1.98 \\
0&0&0&0&0&0&0&0&0&0&9&0&0&0&0&0&9&0&-47.85 & -2.66 \\
0&0&2&0&1&2&0&0&0&1&2&1&2&1&1&2&1&0&-52.83 & -3.30 \\
7&0&0&0&0&0&9&4&0&0&0&0&0&0&0&0&0&0&-60.71 & -3.04 \\
0&2&2&0&2&0&1&2&5&2&0&0&0&0&1&0&0&1&-49.48 & -2.75 \\
1&2&1&0&1&0&2&1&0&0&0&0&2&1&0&0&1&0&-40.21 & -3.35 \\
0&0&16&0&0&0&8&0&8&0&0&8&0&8&0&0&16&0&-185.75 & -2.90 \\
1&2&0&2&1&1&1&3&0&2&0&1&2&1&2&0&0&1&-61.69 & -3.08 \\
0&0&0&0&0&8&0&0&0&0&0&0&0&0&0&0&8&0&-59.34 & -3.71 \\
0&2&0&0&1&0&0&1&2&1&1&0&2&0&1&1&3&1&-42.74 & -2.67 \\
1&0&0&1&0&0&1&3&1&1&2&2&0&0&1&0&1&2&-37.49 & -2.34 \\
3&0&1&0&0&1&0&0&0&1&0&1&1&0&2&2&4&0&-39.57 & -2.47 \\
0&2&0&0&1&2&2&0&0&1&0&2&0&1&0&0&0&1&-33.80 & -2.82 \\
0&0&0&0&1&0&0&0&0&0&0&0&0&0&0&0&15&0&-29.37 & -1.84 \\
0&0&0&0&8&0&0&0&0&8&0&0&0&0&0&0&0&0&-39.13 & -2.45 \\
1&2&0&2&0&0&1&1&1&0&0&2&0&1&0&2&1&2&-45.19 & -2.82 \\
0&0&16&0&16&0&0&0&0&0&0&0&0&16&0&0&0&16&-170.52 & -2.66 \\
0&8&8&8&16&8&24&8&0&8&16&8&0&8&8&0&0&0&-483.13 & -3.77 \\
13&0&0&0&0&7&1&1&0&0&0&0&0&0&0&0&0&0&-88.00 & -4.00 \\
0&1&0&0&2&4&0&0&0&1&2&1&1&3&0&1&0&0&-56.86 & -3.55 \\
9&0&0&0&0&0&0&0&0&0&0&0&0&0&9&0&0&0&-38.97 & -2.17 \\
0&0&0&2&1&0&1&0&1&1&2&0&1&0&1&1&0&1&-33.08 & -2.76 \\
0&1&0&1&0&1&1&0&0&0&0&1&0&0&1&0&0&2&-17.18 & -2.15 \\
0&0&1&1&3&1&0&0&2&0&3&1&1&0&1&1&1&2&-45.47 & -2.53 \\
0&0&0&0&0&0&0&0&0&0&0&0&0&0&9&9&0&0&-45.55 & -2.53 \\
0&0&0&0&1&1&0&0&0&0&0&0&0&2&2&0&0&2&-12.91 & -1.61 \\
0&2&0&0&1&1&1&1&1&1&1&3&0&0&1&1&1&1&-44.00 & -2.75 \\
9&8&9&8&7&10&8&8&4&6&2&7&8&6&5&10&8&5&-370.09 & -2.89 \\
0&0&9&0&0&0&0&0&0&0&0&9&0&0&0&0&0&0&-11.95 & -0.66 \\
1&0&1&1&2&0&0&0&1&0&0&0&0&0&0&1&1&0&-24.14 & -3.02 \\
0&0&0&0&18&0&0&0&0&0&0&0&0&0&0&0&0&2&-83.57 & -4.18 \\
0&0&4&0&0&0&0&0&4&0&0&0&0&0&0&0&0&0&-31.34 & -3.92 \\
0&0&0&0&0&16&0&0&0&0&0&0&32&0&0&0&16&0&-248.64 & -3.88 \\
0&6&0&0&0&0&0&0&0&0&0&0&0&1&0&0&1&0&-3.70 & -0.46 \\
1&1&0&0&0&0&0&0&0&1&1&0&0&1&0&1&0&2&-7.90 & -0.99 \\
1&2&2&0&1&1&2&1&1&0&0&0&0&4&0&3&2&0&-59.84 & -2.99 \\
0&0&1&1&0&0&0&1&1&0&1&0&0&2&1&0&0&0&-27.50 & -3.44 \\
0&2&0&0&1&2&1&1&0&1&1&2&2&2&0&0&1&0&-37.31 & -2.33 \\
0&1&0&0&0&0&0&0&0&0&0&1&0&0&0&6&0&0&-19.17 & -2.40 \\
0&1&0&2&2&2&0&1&1&1&1&0&2&1&1&1&0&0&-56.88 & -3.55 \\
8&0&0&0&0&0&0&8&0&0&0&0&0&0&0&0&0&0&-49.09 & -3.07 \\
0&0&1&0&1&1&2&1&1&2&0&0&2&1&0&1&1&2&-49.73 & -3.11 \\
0&0&2&1&1&0&1&1&0&1&2&1&0&2&1&1&0&2&-42.56 & -2.66 \\
0&0&0&0&0&0&0&18&0&2&0&0&0&0&0&0&0&0&-57.64 & -2.88 \\
1&1&0&0&1&1&1&0&2&3&0&0&2&0&1&1&1&1&-39.10 & -2.44 \\
0&0&0&4&0&0&0&0&0&0&8&0&4&0&0&0&0&0&-18.93 & -1.18 \\
0&0&8&16&16&8&8&16&8&16&0&0&8&16&0&0&0&8&-484.78 & -3.79 \\
8&8&0&0&0&16&0&16&0&0&8&16&16&0&0&16&16&8&-414.13 & -3.24 \\
0&0&0&0&15&0&0&0&0&0&0&1&0&0&0&0&0&0&-81.78 & -5.11 \\
0&2&0&1&0&0&0&2&0&2&0&2&0&2&0&1&3&1&-34.70 & -2.17 \\
1&0&3&1&0&0&1&1&0&1&3&1&0&1&1&0&2&2&-35.55 & -1.98 \\
0&0&1&0&0&0&0&2&0&0&0&3&3&1&2&2&2&0&-50.47 & -3.15 \\
0&0&0&0&0&0&0&0&0&0&0&0&0&0&0&15&1&0&-43.11 & -2.69 \\
0&0&0&32&0&0&0&0&0&0&0&0&0&32&0&0&0&0&-230.94 & -3.61 \\
5&6&6&7&6&6&7&9&5&13&8&5&12&7&3&5&8&10&-326.74 & -2.55 \\
1&0&1&0&2&0&2&2&0&1&4&0&1&1&1&1&1&0&-51.42 & -2.86 \\
0&8&0&16&8&0&0&0&0&0&0&0&8&0&0&8&8&8&-138.38 & -2.16 \\
0&3&0&0&3&3&3&2&2&4&1&0&0&0&6&1&1&3&-68.00 & -2.13 \\
0&0&0&0&0&0&0&0&0&0&0&0&0&32&0&0&32&0&-219.32 & -3.43 \\
0&0&0&0&64&0&0&64&0&0&0&0&0&0&0&0&0&0&-796.72 & -6.22 \\
1&3&0&2&1&1&1&1&3&0&0&0&1&0&1&0&1&0&-51.38 & -3.21 \\
1&0&1&1&1&0&3&0&0&0&0&0&0&1&0&2&0&0&-41.80 & -4.18 \\
0&8&16&0&0&8&0&8&0&0&8&16&0&0&0&0&0&0&-130.89 & -2.05 \\
0&0&0&0&0&0&0&0&8&0&0&0&0&0&0&8&0&0&-45.30 & -2.83 \\
1&0&1&0&0&1&0&0&1&2&2&1&0&0&2&0&2&3&-35.13 & -2.20 \\
0&0&0&0&0&0&0&0&0&10&10&0&0&0&0&0&0&0&-7.75 & -0.39 \\
1&0&0&3&0&0&0&0&1&0&0&1&0&1&0&0&1&0&-14.94 & -1.87 \\
0&8&8&8&8&0&0&0&0&0&0&8&8&8&0&8&0&0&-148.50 & -2.32 \\
0&0&0&0&0&0&0&0&4&0&0&0&0&0&4&0&0&0&-24.20 & -3.02 \\
1&0&0&1&0&0&1&1&1&1&0&1&0&0&0&0&0&1&-13.42 & -1.68 \\
0&0&0&10&0&0&0&6&0&4&0&0&0&0&0&0&0&0&-75.76 & -3.79 \\
0&0&4&0&0&0&0&0&0&0&0&0&4&0&0&0&0&0&-17.00 & -2.13 \\
0&1&1&1&0&0&0&1&0&1&2&1&0&2&1&3&1&1&-42.95 & -2.68 \\
0&0&1&1&1&0&1&1&0&0&1&2&2&1&0&0&0&1&-27.97 & -2.33 \\
		\bottomrule
\end{longtable}
\end{center}

\begin{table}[htbp]
	\caption{The number of atom pairs in the test dataset.}
	\label{tab:app2}
	\centering
    \scalebox{0.85}{
	\begin{tabular}{lllllllllllllllllll}
		\toprule
		 & \multicolumn{1}{c}{H} & \multicolumn{1}{c}{He} & \multicolumn{1}{c}{Li} & \multicolumn{1}{c}{Be} & \multicolumn{1}{c}{B} & \multicolumn{1}{c}{C} & \multicolumn{1}{c}{N} & \multicolumn{1}{c}{O} & \multicolumn{1}{c}{F} & \multicolumn{1}{c}{Ne} & \multicolumn{1}{c}{Na} & \multicolumn{1}{c}{Mg} & \multicolumn{1}{c}{Al} & \multicolumn{1}{c}{Si} & \multicolumn{1}{c}{P} & \multicolumn{1}{c}{S} & \multicolumn{1}{c}{Cl} & \multicolumn{1}{c}{Ar}  \\ \midrule
H&981&403&622&300&277&916&684&1002&564&535&438&327&537&341&741&401&469&372 \\
He&&455&582&554&518&594&601&527&401&637&384&598&586&477&500&780&539&501 \\
Li&&&2091&522&705&543&697&1313&662&676&509&902&741&802&502&432&779&460 \\
Be&&&&1208&671&766&929&590&401&606&414&411&621&1391&283&491&385&442 \\
B&&&&&3324&505&739&3514&659&886&459&474&625&824&572&751&448&551 \\
C&&&&&&1333&681&538&393&554&477&599&841&460&396&868&676&490 \\
N&&&&&&&1068&759&536&927&825&586&525&678&576&433&622&416 \\
O&&&&&&&&2745&387&758&404&520&625&462&505&414&506&455 \\
F&&&&&&&&&815&499&520&369&402&415&488&519&503&372 \\
Ne&&&&&&&&&&1201&717&588&668&621&778&350&498&553 \\
Na&&&&&&&&&&&741&451&374&393&295&346&527&321 \\
Mg&&&&&&&&&&&&917&468&489&302&482&630&383 \\
Al&&&&&&&&&&&&&1515&516&475&599&1106&510 \\
Si&&&&&&&&&&&&&&1098&297&412&889&451 \\
P&&&&&&&&&&&&&&&1155&716&408&342 \\
S&&&&&&&&&&&&&&&&1600&738&464 \\
Cl&&&&&&&&&&&&&&&&&1413&469 \\
Ar&&&&&&&&&&&&&&&&&&421 \\
		\bottomrule
	\end{tabular}
    }
\end{table}
\clearpage



\end{document}